\documentclass[aps,prd,twocolumn,superscriptaddress,nofootinbib,10pt]{revtex4-2}

\usepackage[T1]{fontenc}
\usepackage[utf8]{inputenc}

\usepackage{amsmath,amssymb,mathtools}
\usepackage{tensor}
\usepackage{physics}
\usepackage{lipsum}
\usepackage{bigints}
\usepackage{verbatim}
\usepackage{braket}
\usepackage{cancel}
\usepackage{booktabs}
\usepackage{color}
\usepackage{graphicx}
\usepackage{subcaption}
\usepackage{hyperref}
\usepackage{cleveref}

\crefname{section}{Sec.}{Secs.}
\Crefname{section}{Section}{Sections}

\crefname{appendix}{App.}{Apps.}
\Crefname{appendix}{Appendix}{Appendices}

\newcommand{\Appref}[1]{Appendix~\ref{#1}}

\newcommand{\roughly}[1]{\mathrel{\raise.3ex\hbox{$#1$\kern-0.85em
\lower1ex\hbox{$\sim$}}}}

\newcommand{\mfa}{{\mathfrak a}}

\newcommand{\cO}{{\cal O}}

\newcommand{\MPL}{M_{\rm pl}}

\newcommand{\exd}{{\rm d}}

\newcommand{\nn}{\nonumber}
\newcommand{\rol}{\text{roll}}

\begin{document}

\title{Multi-Field Dilaton Screening Beyond the Thin-Shell Mechanism}

\author{Philippe Brax}
\affiliation{Institut de Physique Th\'eorique, Universit\'e Paris-Saclay, Gif-sur-Yvette, France}

\author{Carsten van de Bruck}
\affiliation{School of Mathematical and Physical Sciences, University of Sheffield, Sheffield, United Kingdom} 

\author{Anne-Christine Davis}
\affiliation{DAMTP \& Kavli Institute of Cosmology (KICC), Cambridge University, Cambridge, United Kingdom}

\author{Adam Smith}
\affiliation{School of Mathematical and Physical Sciences, University of Sheffield, Sheffield, United Kingdom}

\date{\today}

\begin{abstract}
We analyse screening in multi--field scalar--tensor theories, focusing on systems with a dilaton coupled to matter and an axion with a dilaton--dependent kinetic term, in the presence of both planetary and stellar density profiles. Using analytic arguments and fully coupled numerical solutions, we identify a regime in which full screening for a dark--energy--light, effectively unpinned string-dilaton, can occur without fine-tuning. The backreaction of the dilaton's partnered axion field can suppress the exterior scalar charge by selecting a minimum--energy configuration (the BBQ mechanism), yielding robust screening for generic axion gradients. In this regime screening is achieved by cancelling the dilaton's gradient rather than localising it.  This reduces the exterior scalar charge and allows for gravity tests in the solar system to be passed. We then show that the more familiar thin--shell intuition need not apply in the multi-field setting. Axion surface gradients can drastically reshape the dilaton profile and drive a more localised transition without generically suppressing the fifth force. The exterior charge  can remain essentially unchanged or even be enhanced as the shell is made thinner by a kinetically coupled field. Multi--field two--derivative dynamics therefore decouple localisation in thin shells from screening, evade single--field no--go arguments, and reopen viable parameter space for cosmologically light dilaton--like scalars with strong couplings to matter.
\end{abstract}

\maketitle

\section{Introduction}

Modern cosmological data are now precise enough that new light degrees of
freedom are no longer a purely speculative idea. Scalar fields with masses at
or below the Hubble scale are being tested directly through the expansion
history and the growth of structure in the Universe \cite{vandeBruck:2022xbk,Archidiacono:2022iuu,Baryakhtar:2024rky,Baryakhtar:2025uxs,Costa:2025kwt, Castello:2024jmq,Gan:2025nlu,Ghosh:2025pbn, Euclid:2025vml,Zheng:2025vyv,vanderWesthuizen:2025rip}. This has renewed interest in theoretically
controlled scalar sectors that can modify late-time dynamics \cite{Bhattacharya:2024hep,Olguin-Trejo:2018zun,Teixeira:2025czm}, relieve
parameter-level tensions \cite{Wolf:2025jed,Shah:2025ayl,Tiwari:2024gzo,Teixeira:2024qmw,Simon:2024jmu,Zhang:2025dwu}, or imprint signatures that distinguish dynamical dark
energy from a cosmological constant \cite{Bellini:2014fua,Gubitosi:2012hu,Frusciante:2019xia,Efstathiou:2024dvn,Efstathiou:2024xcq,Bedroya:2025fwh,Hong:2025tyi}. In parallel, ultraviolet completions of
gravity such as string theory rarely deliver a low energy effective theory containing a single isolated scalar field  \cite{Bagger:1982fn,Conlon:2006gv,Cicoli:2009zh,Danielsson:2018ztv,Andriot:2025gyr}. They
generically contain multiple light fields, including moduli associated with
couplings and scales and axions protected by approximate shift symmetries,
whose leading interactions are governed as much by field-space geometry as by
potentials \cite{Andriot:2025los}. The appearance of
such scalars in gauge couplings and other fundamental parameters generically
induces couplings to the myriad of fields in the Standard Model. The natural question is therefore not whether light scalars exist,
but how they can remain compatible with stringent local tests while retaining
cosmological relevance.

Any light scalar that couples to matter mediates a new long-range interaction.
Laboratory experiments, solar-system dynamics, and astrophysical observations
place extremely strong bounds on such fifth forces \cite{Kapner:2006si, Bertotti:2003rm, Jain:2012tn, MICROSCOPE:2022doy,Desmond:2018kdn,Naik:2019moz}. Viable scalar-tensor 
theories must therefore include a screening mechanism that suppresses the 
scalar response of dense objects, allowing the field to be active on large 
scales while effectively decoupling in local environments where constraints 
are tightest.

Among the most theoretically well-motivated light scalars, dilatons appear both
as the scalar partner of the graviton in the universal closed-string sector and
its low-energy effective action \cite{Scherk:1974ca,Callan:1985ia,Fradkin:1984pq,Gasperini:1992em},
and as the scalar mode exposed when Jordan--Brans--Dicke--type scalar--tensor
theories are rewritten in the Einstein frame by a conformal rescaling
\cite{Brans:1961sx,PhysRevD.1.3209}; their couplings can be dynamically suppressed by the least--coupling principle
and its runaway and environmentally dependent extensions \cite{Damour:1994zq,Damour:2002nv,Damour:2002mi,Damour:2010rp,Brax:2010gi,Brax:2011ja,Brax:2012gr,Brax:2025ahm}. They provide a minimal parametrisation of how effective 
couplings and masses can depend on a light field. Screening a dilaton with 
constant matter coupling is fundamentally difficult. In the standard telling of the story the dilaton develops a 
density-dependent effective potential and sits at different minima inside and 
outside dense objects. For exponential couplings the field excursion between 
these minima is logarithmic in the density contrast. When considering realistic astrophysical 
density contrasts this excursion is parametrically large, and the dilaton is forced to have non-zero spatial gradients exterior to sources to reach the asymptotic minimum and standard thin-shell 
screening fails \cite{Damour:1994zq,Brax:2010gi}. The fifth force remains essentially unscreened unless the 
coupling is tuned extremely small. This obstruction becomes especially 
constraining in scenarios where cosmological analyses prefer, or comfortably 
accommodate, matter couplings far larger than would be allowed without screening \cite{DiValentino:2017iww,Giare:2024ytc,Smith:2025uaq,Schoneberg:2024ynd}, 
as can occur when a light scalar affects both gravity and microphysics through 
controlled variations of particle masses across key epochs \cite{Damour:1994zq}.

A central point of this paper is that the single-field assumption when drawing conclusions about matter couplings and their correspondence with observables is not
generic from either an ultraviolet or an effective-field-theory viewpoint.
Once more than one light scalar is present, the kinetic sector defines a
curved target-space geometry and associated two-derivative interactions cannot be removed by field redefinitions. In static
configurations, gradients of one scalar can then source another even if
potentials are shallow or symmetries suppress direct couplings. Screening can
therefore be controlled by genuinely multi-field two-derivative dynamics that
do not exist in any one-field truncation. The sign and
magnitude of field-space curvature couplings can decide whether an additional
field helps screening or makes it worse, independently of the detailed form of
the potentials.

In this work we analyse these effects in a minimal axio--dilaton system which has been studied in cosmological settings in various forms in \cite{Catena:2007jf,Burgess:2021obw,Kallosh:2022vha,Alexander:2022own,Bernardo:2022ztc,Smith:2024ayu,Rahimy:2025iyj,Smith:2025grk,Smith:2025ibv,Toomey:2025yuy}, consisting of a dilaton $\chi$ conformally coupled to matter and its partnered axion field $\mfa$ whose kinetic term carries a $\chi$--dependent prefactor $W^2(\chi)$. The resulting curved field space induces a characteristic two--derivative interaction, proportional to $W(\chi)W_{,\chi}(\chi)\,\mfa'(r)^2$, which acts as a source term in the dilaton equation of motion. As a result, axion gradients generically backreact on the dilaton profile even in the absence of direct potential couplings.


The physical effect on the dilaton's evolution depends largely on the sign of $W_{,\chi}$. Given that the dilaton always rolls asymptotically to infinity in the vacuum exterior to a body, if positive, the axion gradient further sources the dilaton and induces a localised dilaton profile wherever it is present. This modifies \emph{where} the dilaton can evolve appreciably and so gives rise to very different profiles compared to where the dilaton would evolve naturally due to its matter coupling alone. In the vacuum exterior to astrophysical bodies a light dilaton will always evolve radially along a Yukawa tail towards its environmental value $\chi_{\rm env}$ according to
\begin{equation}\label{eq:light_dilaton_yukawa_tail}
    \chi\simeq \chi_{\rm env}-\frac{L}{r},
\end{equation} 
where the parameter $L$ is the scalar charge of the dilaton and its size controls the overall exterior spatial gradients associated with the field.
We seek to answer  if this external kinetic driving can generically reduce the dilaton's scalar charge exterior to objects (as reducing the scalar charge amounts to screening the dilaton), e.g. by displacing where the dilaton's evolution occurs, driving the dilaton between its asymptotic minima in a smaller radius (i.e. creating a thinner dilaton shell). We find that the kinetic driving cannot suppress the net scalar charge in any generic regime because of the additional charge penalty associated with adding an externally driven dilaton gradient to the system. However, if $W_{,\chi}$ is negative, the axion gradient generically acts to oppose the dilaton's single field motion towards large $\chi$, and has been shown to reduce generically the dilaton's gradient and effective coupling at the surface of sources\footnote{We shall use the shorthand BBQ for this procedure.} \cite{Brax:2023qyp}. This occurs naturally when combining this effect with a global energy minimisation of the system, which is dynamically selected when the dilaton's surface and asymptotic value are free to vary, removing the need for fine-tuning.

We consider different case studies to illustrate the screening mechanisms in action in this paper, focusing on different variations of stringy dilatons for which the density--dependent effective potential admits minima. Whether local systems are pinned to these minima depends on the in--medium mass scale, and for a dark-energy dilaton the field typically remains unpinned on stellar and planetary scales. In this regime, attempts to screen by forcing a localised interpolation with an external field, i.e. a thin shell, are generically counterproductive, since displacing $\chi$ further from its environmental value without the possibility of reaching an asymptotic stationary point increases the total dilaton excursion and therefore the exterior scalar charge. Viable screening instead proceeds through dynamical gradient suppression when $W_{,\chi}<0$, selected by the minimum--energy configuration of the coupled axio--dilaton system.

If the dilaton is sufficiently massive, or if the system is
sufficiently dense, local objects become pinned to the in--medium minimum. In this case the gradient--suppression
mechanism is generically frustrated from reaching the
global equilibrium configuration. Reducing the exterior
dilaton gradient using an additional field would then require displacing $\chi$ away from its minimum in some interior region, incurring an additional potential-energy cost.
As a result, substantial screening in pinned environments
depends sensitively on environmental details and typically requires tuning of the axion profile or model parameters rather than arising robustly from energy minimisation. Even when the axion kinetic coupling has the opposite sign $W_{,\chi}>0$ and assists rather than opposes the dilaton's roll towards its exterior minimum, the resulting localisation of the transition into a thin shell does not suppress the fifth force. The axion gradient contributes its own charge penalty through the flux balance, so that reducing the shell width $\Delta R/R_s$ leaves the exterior charge $L$ essentially unchanged, decoupling thin--shell formation from actual screening.

These results show that multi--field axio--dilaton dynamics qualitatively alter the connection between profile localisation and observable fifth forces, and motivate treating pinned and unpinned environments separately even within a fixed microscopic model. This paper develops the corresponding analytic criteria and validates them against fully coupled numerical solutions\footnote{The code used in this work is publicly available at
\cite{SmithCode2026}.}, identifying the regions of parameter space in which axio--dilaton models remain compatible with local constraints while retaining cosmological relevance. Surprisingly, we find that screening the dilaton in a multi-field context depends crucially on the sign of the kinetic contribution of the axion to the dilaton's evolution. When negative, screening can be  achieved when the dilaton's  value inside matter is not pinned by its potential. Screening becomes a global property resulting from the minimisation of the scalar field energy. In this case too, screening for pinned fields  follows the same pattern as in the single-field case, i.e.  with a thin shell. This  requires a tuned cancellation between the axion and dilaton contributions to the scalar charge.  This has to be contrasted with the unpinned  case where the dilaton adapts its value at the surface of a compact object and minimises the energy of the field configuration. On the contrary, when the sign of the kinetic coupling between the axion and the dilaton is positive, the minimisation procedure fails and the axion plays hardly any role in the thin shell mechanism. 

The remainder of the paper is organised as follows. \Cref{sec:spherical-setup}
introduces the general spherically symmetric multi--field setup and the scalar
charge that controls fifth forces. 
\Cref{sec:axio-dilaton_setup} then defines the axio--dilaton system and its
two--derivative field--space interactions. \Cref{sec:string_dilatons} introduces the string dilaton and its model features, and then reviews the
single--field dilaton and why thin--shell screening fails for constant coupling. \Cref{sec:thin-shell-piecewise-linear} studies multi--field thin shells and shows that axion--driven localisation does not generically suppress the exterior scalar charge, i.e. screening requires a tuned cancellation between dilaton and axion contributions to the scalar charge.
\Cref{sec:axio-dilaton_no_minimum} presents gradient suppression via energy minimisation (the BBQ mechanism) and identifies when it yields robust charge suppression, including in runaway/no--minimum cases. We revisit the implications for viable dilaton models on astrophysical scales in \Cref{sec:dilaton_revisited} and finally
conclude in \Cref{sec:conclusions}.

\section{Spherically Symmetric Systems}\label{sec:spherical-setup}

\subsection*{General setup and motivation}

We begin with a general scalar--tensor theory in the Einstein frame (minimally coupled to the Ricci scalar $R$),
containing $N$ scalar fields $\phi^I$ with target-space metric $G_{IJ}(\phi^K)$,
a potential $V(\phi^I)$, and a conformal coupling $A(\phi^I)$ to matter,
\begin{align}
S &= \frac{1}{2}\int d^4x \,\sqrt{-g}\;
    \Big[ \MPL^2 R - G_{IJ}(\phi^K)\,\partial_\mu \phi^I \partial^\mu \phi^J
    \nn\\
  &\qquad\qquad\qquad- 2 V(\phi^I) \Big] + S_m\!\left[A^2(\phi^I)\,g_{\mu\nu},\psi_m\right],
\label{eq:action_general}
\end{align}
where $\psi_m$ denotes the matter fields and $\MPL$ is the reduced Planck mass.

Much of the literature on scalar--tensor screening focuses on single-field
models, but this focus turns out to be overly restrictive from the point of
view of effective field theory.  As emphasised in
Refs.~\cite{Coleman:1969sm,Burgess:2007pt,Burgess:2009ea,Penco:2020kvy}, the leading interactions that survive at
low energies (once dangerous zero-derivative potential terms are suppressed by symmetries \cite{Coleman:1969sm,Weinberg:1978kz,Pich:2018ltt}) are the two-derivative $\sigma$-model interactions
\begin{equation}
\label{eq:sigmadef}
\mathcal{L}_{2\text{deriv}}
= \frac12  \sqrt{-g}\,
\Big[ \MPL^2 R - G_{IJ}(\phi^K)\,\partial_\mu \phi^I \partial^\mu \phi^J \Big].
\end{equation}  
These interactions are generically expected to compete with the two-derivative
sector of GR at astrophysical and cosmological energies.

Crucially, these kinetic-metric interactions have \emph{no physical effect} when
the target space is one-dimensional: every one-dimensional manifold is locally
flat, so for a single scalar field the metric can always be removed by a field
redefinition.  
This means that single-field screening models accidentally eliminate the entire
class of two-derivative $\sigma$-model interactions.  
In such models all nontrivial dynamics must come from potentials or matter
couplings, with the kinetic sector playing no role.  
From the EFT point of view this is highly nongeneric. Fundamental UV theories
with symmetries built in \emph{predict} that the kinetic geometry is
the dominant source of scalar interactions in the infrared.

For multiple fields the situation changes qualitatively.  
A curved target space necessarily introduces Christoffel symbols
$\Gamma^I_{\;JK}(\phi)$ which appear in the field equations as genuine 
two-derivative forces.  Specialising \cref{eq:action_general} to static,
spherically symmetric configurations $\phi^I = \phi^I(r)$ and varying with respect to $\phi^I(r)$ gives
\begin{equation}
\frac{1}{r^2}\frac{d}{dr}\!\left(r^2 G_{IJ}\phi^{J\prime}\right)
 - \frac12\,\partial_I G_{JK}\,\phi^{J\prime}\phi^{K\prime}
 = \partial_I V + \partial_I A\,\rho_m(r),
\label{eq:EOM_general}
\end{equation}
where a prime denotes $d/dr$ and $\rho_m(r)$ denotes the conserved matter density.  
The geometric term
\begin{equation}
\frac12\,\partial_I G_{JK}\,\phi^{J\prime}\phi^{K\prime},
\end{equation}
acts as an effective source for $\phi^I$ whenever another field has a
spatial gradient.  
This contribution is purely two-derivative and it persists even when the second field enjoys an
exact shift symmetry.  
As a result, multi-field systems can exhibit low-energy behaviour that has no
counterpart in any single-field reduction.

Because UV-complete theories generically contain several light scalar fields,
and because their low-energy interactions are governed by the curved geometry of
$G_{IJ}(\phi^K)$, the dynamics of any one field are almost never isolated in
practice \cite{Douglas:2006es,Baumann:2014nda}.  
Field-space interactions and turning trajectories couple adiabatic and entropic
directions, so gradients or fluctuations of one scalar can generically source,
drag, or obstruct the rolling of another \cite{Gordon:2000hv,GrootNibbelink:2001qt,Senatore:2010wk}.  
Consequently, single-field screening models omit precisely the class of
interactions that fundamental theory says should be present, and are therefore
likely to be qualitatively inaccurate descriptions of well-motivated scalar--tensor 
physics.\footnote{Unless the additional fields are consistently decoupled or integrated out
within a controlled effective theory \cite{Achucarro:2012sm,Shiu:2011qw} }

The next section seeks to illustrate this point explicitly using the case study of axio-dilaton systems, but first we introduce the appropriate machinery for studying these effects within astrophysical settings.

\subsection*{The scalar charge}\label{sec:scalar-charge}

We focus on a single light field $\phi$ whose scalar force on compact objects is calculated below. This scalar interacts with other scalars via the kinetic couplings appearing in the Klein-Gordon equation (\ref{eq:EOM_general}).
For a static, spherically symmetric configuration of this light scalar $\phi=\phi(r)$
in an approximately homogeneous environment, the solution exterior to a dense body of mass $M$ and radius $R_s$ at radii
$r\gtrsim R_s$ takes the usual Yukawa/Laplace form
\begin{equation}
\phi(r)=\phi_{\rm env}-\frac{L_\phi}{r}\,e^{-m_{\rm env}(r-R_s)} ,
\qquad (r\gtrsim R_s),
\label{eq:yukawa_tail_general}
\end{equation}
where $\phi_{\rm env}$ is the environmental value of the field at large distances from the source, which reduces to 
\begin{equation}\label{eq:light_yukawa_tail}
    \phi\simeq \phi_{\rm env}-\frac{L_\phi}{r},
\end{equation} 
when $m_{\rm env}(r-R_s)\ll 1$, where $m_{\rm env}$ is the environmental mass of the scalar.
The constant $L_\phi$, called the scalar charge, is conserved exterior to the body
\begin{equation}\label{eq:charge_Def}
    L_\phi\equiv r^2\phi'|_{r>R_s}.
\end{equation}
It is the amplitude of the exterior
tail and therefore controls the long-range fifth force.
This solution is valid as long as the scalar potential can be neglected in the vicinity of the compact object. This is the case for instance when the scalar field acts as dark energy. 

To connect $L_\phi$ to the interior dynamics, we specialise \cref{eq:EOM_general} to the  canonically normalised
single-field equation of motion for $\phi$ in the Einstein frame,
\begin{align}
\frac{1}{r^2}\frac{d}{dr}\!\left(r^2\phi'\right)=
\partial_\phi V(\phi^I)\;&+\;\partial_\phi A(\phi^I)\,\rho_m(r)
\nn\\&+\;\frac12\,\partial_\phi G_{JK}(\phi^I)\,\phi^{J\prime}\phi^{K\prime}.
\label{eq:eom_singlefield_general}
\end{align}
Integrating \cref{eq:eom_singlefield_general} from $0$ to $\infty$ and using
regularity at the origin ($r^2\phi'\to 0$ as $r\to 0$), together with the exterior
tail \cref{eq:charge_Def}, gives the flux/charge relation
\begin{align}
L_\phi
&=\int_{0}^{\infty}\!dr\,r^2\bigl[ V,_\phi(\phi^I)
+A,_\phi(\phi^I)\,\rho_m(r)
\nn\\&\qquad\qquad\qquad+\frac12\,\partial_\phi G_{JK}(\phi^I)\,\phi^{J\prime}(r)\phi^{K\prime}(r)
\bigr].
\label{eq:charge_integral_multifield}
\end{align}
The exterior charge is therefore an integrated measure of all sources for $\phi$ across the configuration. Potential gradients and matter couplings contribute, as well as $\sigma$-model interactions encoded in $\partial_\phi G_{JK}\,\phi^{J\prime}\phi^{K\prime}$. Cancellations among these terms can suppress $L_\phi$ even when the fields vary non-trivially inside/outside the object.

If $\phi$ is the only scalar that couples to the matter species, matter follows geodesics of $\tilde g_{\mu\nu}=A^2(\phi)g_{\mu\nu}$, so a test body
experiences an acceleration due to the scalar $\phi$ given by 
\begin{equation}
a_\phi = -\nabla\ln A(\phi)= -\frac{\beta_{\rm env}}{\MPL}\,\nabla\phi,
\;\;
\frac{\beta_{\rm env}(\phi_{\rm env})}{\MPL}\equiv \frac{d\ln A}{d\phi}\Biggr|_{\phi_{\rm env}}.
\label{eq:fifthforce_general}
\end{equation}
This is the case, for instance,  when all the fields but $\phi$ are pseudo-Goldstone bosons with no non-derivative interactions to matter.
For a light field on solar-system scales, the exterior tail
\cref{eq:light_yukawa_tail} implies
\begin{equation}
|a_\phi|\simeq \frac{\beta_{\rm env}}{\MPL}\,\frac{L_\phi}{r^2},
\qquad (r\gtrsim R_s),
\end{equation}
so the scalar force falls off as $1/r^2$, just like Newtonian gravity,
\begin{equation}
|a_N|=\frac{G_NM}{r^2}=\frac{\Phi_N(R_s)\,R_s}{r^2},
\qquad \Phi_N(R_s)\equiv \frac{GM}{R_s}.
\end{equation}
It is therefore convenient to parametrise the amplitude of the exterior scalar tail
by introducing an effective coupling $\beta_{\rm eff}$ through
\begin{equation}
\frac{|a_\phi|}{|a_N|}\;\equiv\;2\,\beta_{\rm env}\,\beta_{\rm eff}
\qquad (m_{\rm env} r\ll 1),
\label{eq:betaeff_motivation}
\end{equation}
which, upon using the expressions above, is equivalent to the charge definition
\begin{equation}
L_\phi \equiv 2\,\beta_{\rm eff}\,M_{\rm pl}\,\Phi_N(R_s)\,R_s.
\label{eq:betaeff_def_general}
\end{equation}
With this definition, the unscreened limit corresponds to $\beta_{\rm eff}\simeq \beta_{\rm env}\simeq \beta$,
giving the familiar force ratio $|a_\phi|/|a_N|\simeq 2\beta^2$.

For a given source there is therefore a direct correspondence between a scalar's exterior charge and its effective coupling strength to matter
\begin{equation}
    \frac{L_\phi}{L_0} = \frac{\beta_{\rm eff}}{\beta_{\rm env}},
\end{equation}
where $L_0$ is the unscreened scalar charge. Screening a scalar therefore corresponds to reducing its net exterior charge. We now introduce two methods of suppressing $L_\phi$ in practice.

\subsection*{Screening mechanisms of interest}

\subsubsection{Thin shells}
\label{sec:thin_shells_intro}

Thin--shell screening is the standard way a matter--coupled scalar evades local
fifth--force bounds when its effective potential 
\begin{equation}\label{eq:general_veff}
    V_{\rm eff}(\phi;\rho_m)=V(\phi)+A(\phi)\rho_m,
\end{equation}
admits density--dependent stationary
points \cite{Khoury:2003aq,Khoury:2003rn, Brax:2010gi,Hinterbichler:2010es}.  For a spherical body with interior density $\rho_c$ and  exterior environmental density $\rho_{\rm env}$, the
field sits near an interior value $\phi_c\simeq \phi_{\min}(\rho_c)$ throughout
most of the volume until a rolling radius $R_{\rm roll}$ and transitions to the ambient value
$\phi_{\rm env}\simeq \phi_{\min}(\rho_{\rm env})$ only in a surface layer of
thickness $\Delta R=R_s-R_{\rm roll}$.  When the scalar is light on the exterior 
scales probed by gravitational tests, the profile matches onto the Yukawa tail \cref{eq:light_yukawa_tail}, so the
observable fifth force is controlled by the exterior charge \cref{eq:charge_Def}.

In the thin--shell regime the charge is sourced only by the small fraction of
mass lying in the rolling layer, yielding the familiar scaling
\begin{equation}
\beta_{\rm eff}\;\simeq\;3\,\beta_{\rm env}\,\frac{\Delta R}{R_s},
\qquad
\frac{L}{L_0}\;\simeq\;3\,\frac{\Delta R}{R_s}.
\label{eq:thin_shell_charge_scaling}
\end{equation}
 The usual thin--shell condition therefore amounts to
$\Delta R/R_s\ll1$, which is
reviewed (and shown to fail) for the single--field dilaton in \Cref{sec:single-field-dilaton}.
A key theme of this paper is that this relationship between a geometrically thin shell and the exterior charge is only true in an idealised single-field model. Once two--derivative field--space
interactions contribute to \cref{eq:charge_integral_multifield}, shell localisation and
fifth--force suppression become independent quantities (see
\Cref{sec:thin-shell-piecewise-linear}).

\subsubsection{BBQ Screening}
\label{ss:bbq_intro}

A different route to screening is possible when the shell picture itself is not
well defined, either because $V_{\rm eff}$ has no relevant density--dependent
minima or because the scalar is too light to relax to them on the scale of the
source ($m_{\rm eff}(\rho_c)R_s\ll1$).  In this regime $\phi(r)$ can
roll throughout the interior and one should expect $\Delta R/R_s\sim1$.
Nevertheless, the fifth force can still be suppressed if the static
configuration dynamically selects a surface value that minimises the total
energy, which also happens to reduce the exterior gradient energy $\sim L_\phi^2/R_s$ and hence
the conserved charge.  We refer to this energy--selected charge suppression as
the BBQ mechanism (following the minimisation logic of \cite{Brax:2023qyp}).

In multi--field systems this becomes possible because the gradient energy of one
field can depend on another through the target--space metric $G_{IJ}(\phi)$, and
the same curved field--space structure contributes directly to the charge
integral through the $\sigma$--model term in \cref{eq:charge_integral_multifield}.  For the sign choices in which these
two--derivative interactions can oppose the matter--induced source for the
field of interest, the minimum--energy configuration can dynamically suppress
the exterior charge $L_\phi$ even when no thin-shell forms \cite{Burgess:2021qti}.  This is
the organising principle for the runaway/unpinned analyses in
\Cref{sec:axio-dilaton_no_minimum}.

We now proceed to specifying the system of interest in this paper and begin applying the astrophysical toolkit laid out in this section.

\section{Multi--field axio--dilaton systems}\label{sec:axio-dilaton_setup}

The considerations above suggest studying at least two scalar fields when analysing
screening of fifth forces. Among the simplest and best-motivated examples are
axio--dilaton systems, consisting of an axion $\mfa$ and dilaton $\chi$, specified in terms of the covariant language as
\begin{equation}
  \phi^I=(\chi,\mfa),\qquad 
  G_{IJ}=\mathrm{diag}\!\left(1,\,W^2(\chi)\right).
\end{equation} 
This means the full action for our setup has the form\footnote{Unless explicitly stated otherwise we use fundamental units $c=\hbar=1$ throughout and define the reduced Planck mass in terms of Newton's constant by $\MPL^{-2} = 8\pi \rm G_N$.}
\begin{align}\label{eq:Action}
    S = \!\int \exd^4x \sqrt{-g}
    \Biggl\{
        \frac{\MPL^2}{2}
        \Bigl[
            &R
            - \partial_\mu\chi\,\partial^\mu\chi
            - W^2(\chi)\,\partial_\mu\mfa\,\partial^\mu\mfa
        \Bigr]
        \nn\\&
        - V(\chi,\mfa)
    \Biggr\}
    + S_m(A^2(\chi)\,g_{\mu\nu},\Psi_m)\,,
\end{align}
so that $\chi$ is a canonically normalised dilaton with a universal coupling to the trace of the
energy--momentum tensor (through $A(\chi)$), while the axion $\mfa$ is a shift--symmetric pseudoscalar whose
two--derivative term is multiplied by a dilaton--dependent prefactor $W^2(\chi)$.
In ultraviolet completions where $\chi$ and $\mfa$ form a supersymmetric pair, $W(\chi)$ is
not an arbitrary function but is fixed by the geometry of the scalar manifold, i.e.\ by the
field--space metric inherited from the K\"ahler potential \cite{Polchinski:1998rr,Grimm:2004uq}. A widely used and well--motivated
parameterisation is the exponential form\footnote{We also consider quadratic functional forms for $W^2(\chi)$ in \Cref{app:quadraticW} to show the behaviour of the system and our conclusions remain robust in the presence of different kinetic coupling functional forms.}
\begin{equation}\label{eq:exp_W}
  W(\chi)=W_0 e^{\zeta \chi},
\end{equation}
where $\zeta$ encodes the (constant) curvature scale of the two--dimensional target space \cite{Carrasco:2015uma,Cicoli:2023opf,Apers:2024ffe}. The physical consequence of this curved field space is that axion gradients backreact on the
dilaton through the connection term in the covariant equations of motion, yielding an
effective source for $\chi$.

For spherically symmetric and static configurations the field equations reduce to
\begin{align}
    \mfa'' + \frac{2}{r}\mfa' 
        &= -2\,\frac{W_{,\chi}}{W}\,\chi'\mfa'
           \nn\\&\qquad+ \frac{1}{W^2(\chi)}
             \Big[
                 \partial_{\mfa}V(\chi,\mfa)
                 + \partial_{\mfa}U(\mfa)A(\chi)\,\rho_m(r)
             \Big],
             \label{eq:eom_axion}
             \\[4pt]
    \chi'' + \frac{2}{r}\chi'
        &= W(\chi)W_{,\chi}(\chi)\,\mfa'^2
           \nn\\&\qquad+ \partial_\chi V(\chi,\mfa)
           + \partial_\chi A(\chi)\,[1+U(\mfa)]\rho_m(r).
           \label{eq:eom_dilaton}
\end{align}
 These equations exhibit explicitly the two--derivative interaction induced by the curved field space, where the axion sources $\chi$ only where $\mfa'(r)\neq 0$, and the strength of this source is controlled by the geometry through $W W_{,\chi}$.

The dilaton charge analogue of \cref{eq:charge_integral_multifield} controlling the coupling strength of the dilaton to matter in this system is given by
\begin{equation}
  L = \big[r^2\chi'(r)\big]_{0}^{\infty}
    = \int_0^\infty dr\, r^2\Big[
        W W_{,\chi}\, \mfa'(r)^2
        + V_{,\chi}
        + A_{,\chi}\rho_m
      \Big],
  \label{eq:global-L-def}
\end{equation}
where $\rho_m$ is the matter density. Finding a way to suppress this charge, either by suppressing contributions or identifying natural regimes where they largely cancel is the purpose of this paper.

\subsubsection*{Dilatons in the literature}

To place this model in context, it is useful to recall that ``dilaton''
originally denotes the (pseudo-)Goldstone boson of broken scale invariance
\cite{Wetterich:1987fm,Wetterich:2010kd,Henz:2016aoh,Wetterich:2020cxq,Burgess:2021obw},
while in much of the scalar--tensor and string--moduli literature it is used more broadly,
often as an umbrella term for light moduli whose variations induce an
(approximately) universal conformal rescaling of the matter sector
\cite{Scherk:1974ca,Fradkin:1984pq,Callan:1985ia,Gasperini:2001pc,Flanagan:2004bz,Will:2014kxa}. Across this literature,
the qualitative behaviour is largely dictated by three interrelated pieces of structure:
 the sign and magnitude of the matter coupling $\beta\equiv  \partial\ln A/\partial\chi$,
the slope of the underlying runaway potential for $\chi$ (often approximately
exponential in the Einstein frame), and, in multi--field settings, the sign of the
axion--dilaton kinetic coupling $\zeta$ which fixes the sign of $W_{,\chi}$. These sign choices
control whether the density--dependent effective potential for $\chi$ admits local minima
(pinning the field in different environments) or
instead remains runaway. With this motivation, we summarise in \Cref{tab:Summary_of_cases} the main dilaton
classes studied to date and the sign choices most commonly realised in each construction, for the dilaton's potential 
\begin{equation}\label{eq:exp_pot}
    V = V_0 e^{\lambda\chi/\MPL},
\end{equation}
and matter coupling function 
\begin{equation}\label{eq:exp_matter}
    A(\chi) = e^{\beta\chi/\MPL}.
\end{equation}
This
serves as a map for the regimes analysed in the remainder of the paper.

The \emph{string dilaton} corresponds to the gravi--dilaton actions studied in
\cite{Green:2012oqa, Damour:2002nv, Gasperini:2001pc, Brax:2010gi}. By contrast, \emph{yoga} models
\cite{Burgess:2021obw, Brax:2023qyp, Smith:2024ayu} arise when the axion is the imaginary
partner of the volume modulus $\chi$. Both cases typically predict $\beta\sim\mathcal{O}(0.1)$
couplings to matter.
In yoga models the dilaton potential is engineered to possess a shallow vacuum minimum at the dark--energy scale, so that the field is asymptotically stabilised on cosmological scales. Adding matter shifts this minimum only weakly so that the density--dependent effective potential retains a nearby minimum, but its curvature remains set by the dark--energy scale and is therefore far too small to pin the field in astrophysical environments. So although the dilaton is cosmologically stabilised it behaves as effectively unpinned around compact objects and a weakly fixed asymptotic value determined cosmologically. 

\begin{table*}
\centering
\makebox[0.9\textwidth][c]{%
\begin{tabular}{l @{\hskip 1.2em} c @{\hskip 1.2em} c @{\hskip 1.2em} c
@{\hskip 1.2em} c @{\hskip 1.2em} c}
\hline\hline
\strut Model  & \strut $\beta$ & \strut $\lambda$
& \shortstack{\strut Thin Shell\\[-0.2ex]($W_{,\chi} > 0$)}
& \shortstack{\strut BBQ, unpinned\\[-0.2ex]($W_{,\chi} < 0$, $m_{\rm eff}R_s \ll 1$)}
& \shortstack{\strut BBQ, pinned\\[-0.2ex]($W_{,\chi} < 0$, $m_{\rm eff}R_s \gg 1$)}\\
\midrule
String Dilaton & $> 0$   & $< 0$  & No & Yes & Environmentally fine-tuned\\
Yoga           & $< 0$   & $< 0$  & No & Yes & --\\
\bottomrule
\end{tabular}
}
\caption{Summary of different dilaton models and the signs of their couplings to the standard model particles, $\beta$, and potential slopes, $\lambda$, given in \cref{eq:exp_pot,eq:exp_matter}. Dilatons with opposite sign $\beta$ and $\lambda$ exhibit minima in their effective potentials, while those with identical signs do not. We distinguish the BBQ regime ($W_{,\chi}<0$) into an \emph{unpinned} case ($m_{\rm eff}R_s\ll 1$) and a \emph{pinned} case ($m_{\rm eff}R_s\gg 1$).}
\label{tab:Summary_of_cases}
\end{table*}

\subsubsection*{Axions in the literature}

The axion sector contains two contributions: a joint potential $V(\chi,\mfa)$, typically periodic in the axion direction \cite{Marsh:2015xka} as in ordinary axion models\footnote{Although axions with exact quadratic potentials can be found in dual axion frameworks \cite{Burgess:2025geh}}, and a matter--induced coupling function $U(\mfa)$ arising from the coupling to the Einstein-frame matter energy density 
\begin{equation}
    \rho_{m\,E} = A(\chi)(1+U(\mfa))\rho_m(r),
\end{equation}
where $\rho_m$ is the conserved mass density and in what follows we assume $U(\mfa) \ll 1$. The two fields roll in their respective directions of the effective potential
\begin{equation}\label{eq:full_exponential_potential}
    V_{\rm eff}(\chi,\mfa;\rho_m) = V(\chi,\mfa) + [1+U(\mfa)]\,\rho_m(r)A(\chi),
\end{equation}
 whose density–dependent stationary points determine the asymptotic interior and exterior values of $\mfa$ in regions of approximately constant density.  The interior and exterior equilibrium values
\begin{equation}
    \mfa_- \equiv \mfa(r=0), \qquad
    \mfa_+ \equiv \mfa(r\gg R_s),
\end{equation}
are obtained from the local force--balance relation $\partial_{\mfa}V_{\rm eff}=0$ evaluated at the interior density $\rho_c$ and the exterior density $\rho_{\rm env}$.  These definitions rely only on the existence of local stationary points and do not assume any particular microphysical form for $V$ or $U$. 
In general the axion force-balance condition defines a density-dependent valley $\,\mfa_\star(\chi;\rho)\,$ in field space. 

The thin--shell estimates below assume that the axion transition is much more localised than the dilaton transition.
This implies the axion should be heavy on the scale of the object so that it interpolates only in a narrow surface band,
\begin{equation}
\ell_{\mfa}\sim m_{\mfa,{\rm sh}}^{-1}\ll \Delta R\lesssim R_s\,,
\end{equation}
where $m_{\mfa,{\rm sh}}$ is the axion effective mass in the surface/transition region. In this regime $\mfa'(r)$ has support
only very close to $r\simeq R_s$, so the axion backreaction term acts as a near--surface kick in the dilaton equation rather than an extended
source.

\begin{figure}
    \centering
    \includegraphics[width=\linewidth]{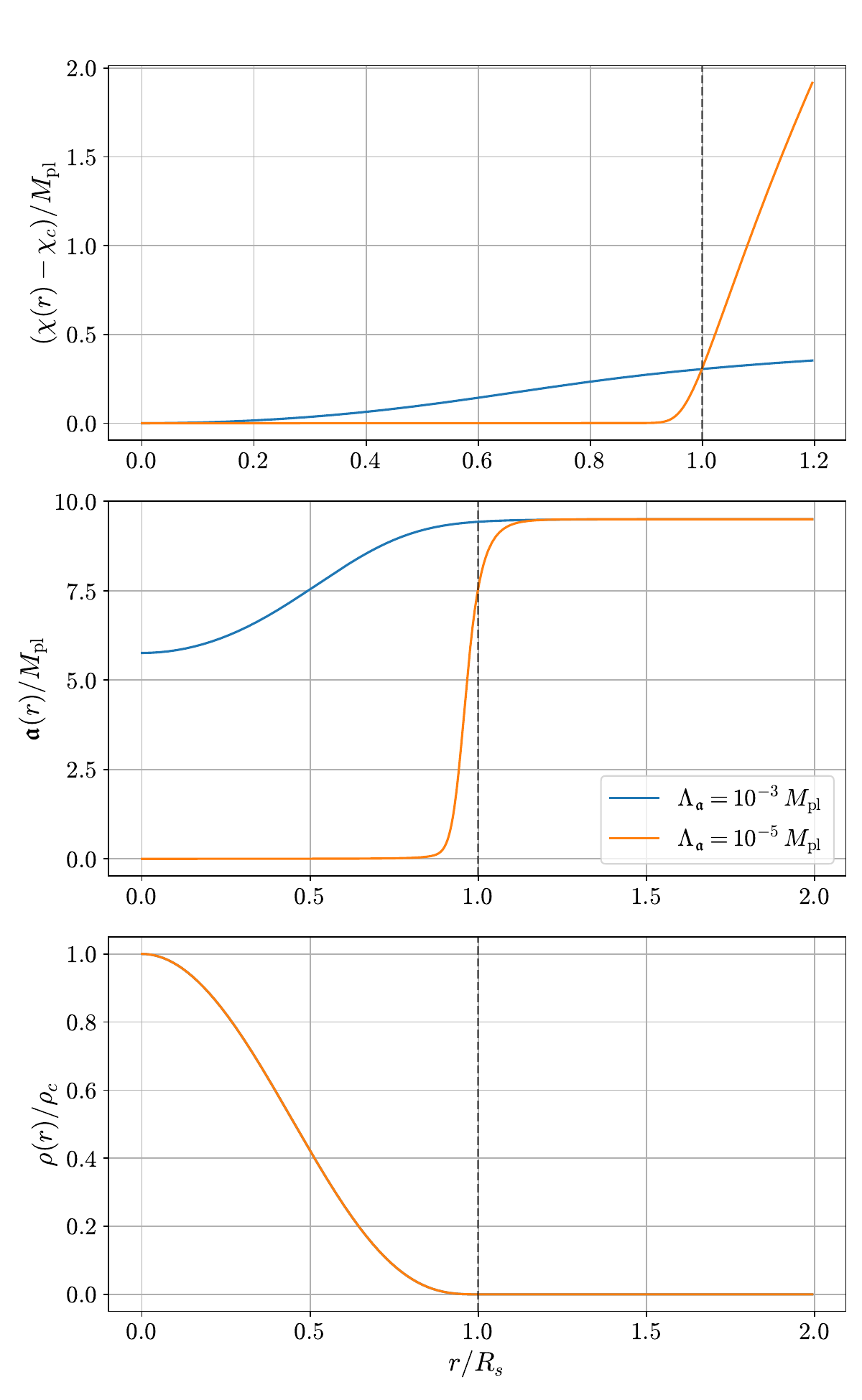}
    \caption{Axion--dilaton profiles illustrating the scale hierarchy underlying the thin--shell approximation. 
\textbf{Top:} dilaton displacement; \textbf{middle:} axion; \textbf{bottom:} conserved density profile.
The dashed line marks the surface $r=R_s$. The two curves correspond to the cut-off scale in \cref{eq:U_ax} being $\Lambda_\mfa=10^{-3}\MPL$ (broad axion transition)
and $\Lambda_\mfa=10^{-5}\MPL$ (surface--localised axion transition), showing how a macroscopic axion gradient drives $\chi$ well
inside the object. In both cases, $W_0 = 1$, $\zeta = \sqrt{2}$, and the vacuum axion mass is taken to be $m_\mfa = 5\times10^{-17}\rm eV$.}
    \label{fig:light_axion}
\end{figure}

This scale hierarchy is illustrated in \Cref{fig:light_axion}. In the middle panel, the case corresponding to a heavier axion inside the source (orange) shows an axion interpolation confined to a thin layer near the surface, whereas for a lighter axion (blue) the axion varies over $\mathcal{O}(R_s)$ and its gradients persist well away from $r\simeq R_s$.
The top panel shows the consequence for the dilaton: when the axion gradient is localised $\chi$ remains close to
its interior value until near the surface, but when the axion is light ($m_{a,{\rm sh}}R_s\lesssim 1$,
blue) the same two--derivative coupling drives $\chi$ off its interior plateau well before $r\simeq R_s$. The usual
thin--shell assumption that $\chi$ stays approximately stationary until a radius close to the surface is therefore not
self--consistent in this regime.\footnote{We treat the breakdown of the mass hierarchy between the axion and dilaton explicitly in the fully coupled numerical
solutions of \Cref{sec:thin-shell-piecewise-linear}.}

When the axion is heavy on macroscopic scales one may integrate it out. For fixed $(\chi,\rho_m)$ the axion rapidly relaxes to $\mfa_\star(\chi;\rho_m)$, defined by
\begin{equation}\label{eq:force_balance}
    \partial_{\mfa}V_{\rm eff}\!\left(\chi,\mfa;\rho_m\right)\big|_{\mfa=\mfa_\star}=0,
\end{equation}
which yields a reduced one--field effective potential
\begin{equation}\label{eq:reduced_potential}
    V_{\rm eff}^{\rm red}(\chi;\rho_m)\equiv V_{\rm eff}\!\left(\chi,\mfa_\star(\chi;\rho_m);\rho_m\right),
\end{equation}
for the light dilaton to evolve in. This forms the basis of the analytic investigations conducted in \Cref{sec:thin-shell-piecewise-linear,sec:axio-dilaton_no_minimum}

\subsubsection*{Piecewise linear axion profile}

To determine how an axion gradient modifies the usual dilaton profile in the analytic approximations used below, we adopt a representative piecewise linear axion profile. This can be seen as a leading-order Taylor expansion of the axion gradient around a very steep transition localised close to the surface of the object. The axion must interpolate between its interior and exterior equilibrium values set by the density--dependent effective potential \cref{eq:full_exponential_potential}, and for understanding its effects on other light scalars, all that matters is the total jump $\Delta\mfa=\mfa_+-\mfa_-$ and the width $\ell_\mfa$ of the region in which $\mfa'(r)$ is nonzero. This motivates the effective ramp
\begin{equation}
    \mfa(r)=
    \begin{cases}
        \mfa_- , & r < \alpha R_s, \\[4pt]
        \mfa_- + (\Delta \mfa/\ell_\mfa)(r - \alpha R_s), & \alpha R_s < r < R_s, \\[4pt]
        \mfa_+ , & r > R_s ,
    \end{cases}
    \label{eq:axion_ramp_final}
\end{equation}
with $\ell_\mfa=(1-\alpha)R_s$ and $\alpha\in[0,1]$. Here $\ell_\mfa$ is the characteristic width of the ramp, which would correspond to the inverse Compton wavelength of the axion, $\lambda_\mfa\sim 1/m_{\mfa\,\text{in}}$, in the transition region between the interior and exterior of the object, where $m_{\mfa\,\text{in}}(r)$ is the interior mass of the axion. 

In practice, axion profiles of this type can be realised using relatively
simple microphysical potentials. In the full two--field numerical analysis
presented below, we adopt separable forms for the scalar potentials,
\begin{equation}
    V(\chi,\mfa) = V_\chi(\chi) + V_\mfa(\mfa),
\end{equation}
with an axion vacuum potential
\begin{equation}\label{eq:V_ax}
    V_\mfa(\mfa) = \frac{1}{2} m_\mfa^2 \left(\mfa - \mfa_+\right)^2,
\end{equation}
and a matter--induced coupling function
\begin{equation}\label{eq:U_ax}
    U(\mfa) = \frac{1}{2}\frac{\left(\mfa - \mfa_-\right)^2}{\Lambda_\mfa^2}.
\end{equation}
In this setup, the vacuum contribution favours the minimum at $\mfa_+$,
while the matter coupling shifts the effective minimum towards $\mfa_-$ in
regions of sufficiently high density. The axion therefore interpolates
between $\mfa_+$ and $\mfa_-$ depending on whether the vacuum or
matter contributions dominate the effective potential.
The detailed shape and width of this transition are controlled by the
microscopic vacuum mass $m_\mfa$ and the matter--coupling scale
$\Lambda_\mfa$. These functions can be viewed as leading order Taylor expansions around minima of the potentials.

Having introduced generic axio-dilaton models and the phenomenology of the respective fields, we now dive into the parameter choices and models specifics for string dilatons.

\section{String Dilatons}\label{sec:string_dilatons}

\subsection{The models}

The purpose of this section is to analyse the case of a stringy dilaton, by which we refer to  the
environmentally dependent (runaway) dilaton scenarios of
\cite{Damour:1994zq,Brax:2010gi} (see also the screening review
\cite{Burrage:2017qrf}).  In the Einstein frame, a conformal coupling
$\tilde g_{\mu\nu}=A^2(\chi)\,g_{\mu\nu}$ introduced in \cref{eq:Action} implies that non-relativistic matter
acts as a $\chi$-dependent source, so static configurations can be organised by
an effective potential composed of both matter and vacuum contributions \cite{Burrage:2017qrf}, as in \cref{eq:general_veff}.  For the toy
constant-coupling choice
\begin{equation}
  A(\chi)=e^{\beta\chi/\MPL},\qquad \beta=\text{const},
\end{equation}
and for a runaway dilaton potential of the form used in \cite{Brax:2010gi},
$V(\chi)\propto A^4(\chi)\,e^{-\chi/\MPL}$, the Einstein-frame self-interaction
reduces to an exponential,
\begin{equation}
  V_\chi(\chi)=V_0\,e^{(4\beta-1)\chi/\MPL},
\end{equation}
so that
\begin{equation}\label{eq:single_field_eff_pot}
  V_{\rm eff\,\chi}(\chi;\rho_m)=V_\chi(\chi)+A(\chi)\,\rho_m(r),
\end{equation}
with $\rho_m$ the conserved Einstein-frame matter density.
For reference, in \cite{Brax:2010gi} the cosmological value of the coupling today
is $\beta_{\rm cos}\simeq 0.23$, so throughout we take $\beta=0.2$ as a representative coupling before screening effects are included.

Because $\beta=0.2$ implies $A(\chi)$ is growing, while $V(\chi)$ is shrinking,
minimising $V_{\rm eff\,\chi}$ yields two equilibrium values, $\chi_c$ and $\chi_{\rm env}$, corresponding to the core and exterior densities. A key subtlety is that the interior value of the dilaton is \emph{not}
guaranteed to ever reach the dense--matter minimum $\chi_c$ in all dense objects. Whether the field can relax to $\chi_c$ within a finite
object depends on the local curvature of the effective potential. 

The pinned regime requires the effective mass, defined as
\begin{equation}
    m_{\rm eff}(\rho_m) \equiv \frac{\partial^2 V_{\rm eff\,\chi}}{\partial\chi^2},
\end{equation}
to satisfy
\begin{equation}
    \text{Pinned} :\qquad m_{\rm eff}(\rho_c)R_s\gtrsim 1.
\end{equation}
For an ultralight dilaton with 
\begin{equation}
    \text{Unpinned} :\qquad m_{\rm eff}(\rho_c)R_s\ll 1,
\end{equation}
the field
cannot track $\chi_c$ and remains close to $\chi_{\rm env}$ throughout most of
the interior, up to $\mathcal{O}(\Phi_N)$ perturbations. Numerically, $m_{\rm eff}(\rho_c)R_s\gtrsim 1$ corresponds to $m_{\rm eff}\gtrsim
2.8\times 10^{-16}\,{\rm eV}$ for the Sun ($R_\odot=7.0\times 10^8\,{\rm m}$),
and $m_{\rm eff}\gtrsim 3.1\times 10^{-14}\,{\rm eV}$ for the Earth
($R_\oplus=6.4\times 10^6\,{\rm m}$), showing they do not pin the dilaton if it is cosmologically light.\footnote{Neutron stars are an important exception: these stars are many orders of magnitude more dense, making the effective potential for the dilaton more steep, and one can have $m_{\rm eff}(\rho_{\rm NS})R_{\rm NS}\gtrsim1$ even when the vacuum mass is cosmological. The field is then pinned in the interior and develops a thin--shell--type profile in a relativistic stellar background; see e.g.\ \cite{Tsujikawa:2009yf, Brax:2017wcj}.
}  By contrast, a dark-energy-like mass
$m_{\rm eff}\sim H_0\sim 10^{-33}\,{\rm eV}$ gives $m_{\rm eff}R_\odot\sim 10^{-17}\ll 1$. 

Pinning for cosmologically light dilatons can nevertheless be realised in models
where the effective potential is sufficiently steep in dense regions, so that
the local minimum shifts substantially between vacuum and matter and the
curvature $m_{\rm eff}^2(\rho)=\partial_\chi^2V_{\rm eff}$ increases rapidly
with density.  In such scenarios the dilaton can remain light on cosmological
scales while becoming comparatively heavy in high-density environments, and the
appropriate screening mechanism is then selected dynamically by the local value
of $m_{\rm eff}(\rho)$.

A useful way to quantify the required potential steepness in dense
regions is via the scaling of the curvature at the density-dependent minimum,
$m_{\rm eff}(\rho)\propto \rho^{p}$.  Pinning on a length scale $\lambda$ requires
$m_{\rm eff}(\rho)\gtrsim \lambda^{-1}$, and between two environments one has
$m_{\rm eff}(\rho_2)/m_{\rm eff}(\rho_1)=(\rho_2/\rho_1)^p$.  Given the enormous
density contrast between cosmological and stellar/planetary environments, even
moderate $p$ can yield many orders-of-magnitude increases in $m_{\rm eff}$.  For
example, $\rho_\star/\rho_{\rm cos}\sim 10^{29}$ implies
$m_\star/m_{\rm cos}\sim 10^{10}$ for $p=1/3$, while $p\simeq 1/2$--$2/3$ gives
$m_\star/m_{\rm cos}\sim 10^{15}$--$10^{20}$.  In standard chameleon potentials
$V\propto \chi^{-n}$ one finds
$m_{\rm eff}(\rho)\propto\rho^{(n+2)/[2(n+1)]}$ \cite{Burrage:2017qrf}. For the canonical chameleon case $n=1$ with the potential scale fixed to the
dark-energy value, recent laboratory (e.g. \cite{Yin:2022geb}) and fifth-force searches place extremely
strong constraints, excluding essentially the entire cosmologically viable
parameter space across a wide range of $\beta$.
Larger values of $n$ generally remain viable, but within restricted regions
of the $(\beta,n)$ parameter space \cite{Burrage:2016bwy, Fischer:2024eic}. This means that pinning a cosmologically light scalar on Solar--System scales requires an exceptionally rapid growth of $m_{\rm eff}$ with density, leaving only small corners of chameleon-like parameter space (although increasing $n$ does reopen viable regions).

Another important point to note about the upcoming analysis is the preceding discussion treats the environment through a single
background value $\chi_{\rm env}$ and thus implicitly assumes a separation of
scales between compact objects and the larger-scale host potential.  In the
presence of strong non-linearities this need not be a reliable assumption. Overlapping non-linear regions leading to multi-source boundary
conditions, and spatially varying screening efficiencies can modify the
effective ambient value and the resulting exterior charge.  A definitive
assessment of the galactic/solar environment therefore requires either a
dedicated many-body treatment or cosmological/galactic N-body simulations that
resolve the coupled scalar dynamics.\footnote{
Nonlinear screening in scalar--tensor theories has been widely studied using
modified-gravity $N$-body simulations that evolve the scalar field alongside
the matter distribution; see for example
\cite{Oyaizu:2008sr,Oyaizu:2008tb,Li_2012,Brax_2012,Winther:2014cia,Winther:2015wla}.
These simulations primarily focus on large-scale structure and dark-matter halo
formation on cosmological scales, while simulations resolving the scalar field
profiles of stellar-scale compact objects embedded in realistic galactic
environments remain largely unexplored.
}  We do not attempt such a calculation
here, and instead adopt $\chi_{\rm env}$ as an effective parameter encoding the
local environment.

With all of these complications and caveats in mind we use the rest of this paper to see what scenarios work in practice with the screening mechanisms at hand for these types of scalars.

\subsection{Thin-shell failure for a single field dilaton}
\label{sec:single-field-dilaton}

Before analysing the full axio--dilaton system, it is useful to recall the
standard behaviour of a single dilaton field coupled to matter. 

In order to model the dynamics of the single field dilaton in the solar system, e.g.\ around a planet, we consider the dynamics of this two--field system in the presence of a uniform--density, spherically symmetric source of radius $R_s$ and density $\rho_m(r<R_s)=\rho_c$, embedded in an exterior environment with density $\rho_m(r>R_s)=\rho_{\rm env}$. If pinned, the dilaton will sit at the interior and exterior equilibrium values $\chi_c$ and $\chi_{\rm env}$ of \cref{eq:single_field_eff_pot} below and above the surface of the source, and for a thin shell to form, the field must interpolate between these values close to the source's surface, as discussed in \Cref{sec:thin_shells_intro}. Following the prescription in \cite{Brax:2010gi}, for a spherical body with surface Newtonian potential 
$\Phi_N(R_s)$, the standard thin--shell condition can be evaluated as
\begin{equation}
    \frac{\Delta R}{R_s}
    = \frac{|\chi_{\rm env} - \chi_c|/\MPL}
           {3\,\beta\,\Phi_N(R_s)}
    \ll 1,
    \label{eq:singlefield_thinshell}
\end{equation}
where $\Delta R = R_s - R_{\rm roll}$ is the thickness of the layer in which the
field departs from the interior value.

For the exponential coupling the field excursion evaluates to the well-known form
\begin{equation}
    |\chi_{\rm env} - \chi_c |
    \simeq
    \frac{\MPL}{1-3\beta}
    \left|\ln\!\frac{\rho_{\rm env}}{\rho_c}\right|.
    \label{eq:singlefield_excursion}
\end{equation}
For any realistic density contrast, this excursion is extremely large.
Since the denominator of \cref{eq:singlefield_thinshell} is tiny for astrophysical
objects ($\Phi_N \sim 10^{-6}$ for the Sun), the ratio in \cref{eq:singlefield_thinshell} satisfies
\begin{equation}
\frac{\Delta R}{R_s} \gg \mathcal{O}(1),
\end{equation}
and no thin shell forms.

As a consequence of \cref{eq:thin_shell_charge_scaling}, the scalar force mediated by $\chi$ remains unscreened and has
strength $\sim\beta^2$ relative to gravity.  Compatibility with solar--system
tests therefore requires $\beta \lesssim  10^{-3}$, making the single--field dilaton
effectively inert.  
This constitutes the robust no--go theorem for dilaton screening with constant
coupling~\cite{Brax:2010kv}.

The failure of the single--field dilaton is not unsurprising. As emphasised in \Cref{sec:spherical-setup}, single-field models remove the geometric
$\sigma$-model interactions that generically dominate multi-field low-energy
dynamics. We now show explicitly how the presence of a second heavy
field (an axion with a field-dependent kinetic prefactor) modifies the shell
balance and qualitatively alters the screening behaviour.

\section{Multi-field thin-shells (with $W_{,\chi}>0$)}
\label{sec:thin-shell-piecewise-linear}

\subsection{Axions and their effects}
We now turn to accounting for the deviations in the shell profile and charge-shell radius relation in the presence of the dilaton's partnered axion field. In this case, the interior and exterior dilaton values $\chi_c$ and $\chi_{\rm env}$ are minima of the reduced effective potential \cref{eq:reduced_potential}, which depends on the equilibrium values $\mfa_+$ and $\mfa_-$, with $\mfa_\pm$ determined by the force--balance condition \cref{eq:force_balance} and the full effective potential \cref{eq:full_exponential_potential}. 

\medskip

As discussed in \Cref{sec:axio-dilaton_setup}, in the parameter regime of analytical interest here the axion is heavier than the dilaton on macroscopic scales, so that the ramp is localised in a narrow region close to the surface,
\begin{equation}
\ell_\mfa \ll \Delta R\,.
\end{equation}
Consequently, the axion gradient contributes to the dilaton equation of motion only through a near--surface source localised within the ramp, and its effect is controlled by the ramp data $(\Delta \mfa,\ell_\mfa)$. In this region, the axion gradient gives the dilaton a near--surface kick in $\chi'$ whose direction is determined by $W W_{,\chi}$, as seen in \cref{eq:eom_dilaton}. We focus to start with on the case of $W_{,\chi}\chi'(r)>0$, corresponding to the axion driving being the same sign as the dilaton's velocity, further accelerating the dilaton and driving it between its minima more efficiently, in turn reducing the dilaton's shell width.

The multi-field thin-shell analysis in this case reduces to evaluating the
integrated version of \cref{eq:eom_dilaton} across $[R_{\rol},R_s]$. Multiplying \cref{eq:eom_dilaton} by $r^{2}$ and integrating only across the
region where $\chi$ rolls, the left-hand side is a total
derivative equivalent to the dilaton charge,
\begin{equation}\label{eq:L_thin_shell}
L = \int_{R_{\rol}}^{R_s}\!dr\,\frac{d}{dr}\!\big[r^2\chi'(r)\big]
   = \big[r^2\chi'(r)\big]_{R_{\rol}}^{R_s}
   \simeq \frac{R_s^2}{\Delta R}\,\Delta\chi,
\end{equation}
where we used the thin--shell approximation $r\simeq R_s$ throughout the layer and the no-roll condition outside the shell $\chi'(R_{\rm roll})=0$. The integrated
equation yields the global shell balance relation
\begin{align}
\int_{R_{\rol}}^{R_s}\! r^2 V_{,\chi}\,dr
+
\int_{R_{\rol}}^{R_s}\! r^2 A_{,\chi}\rho_m\,dr
+
\int_{R_{\rol}}^{R_s}\! r^2 W W_{,\chi}\,\mfa'^2\,dr \nn \\
=
L .
\label{eq:shell-balance}
\end{align}

We now evaluate each term in the thin--shell regime, where all non-vanishing
support is restricted to narrow intervals near $r\simeq R_s$.

\paragraph{Axion--gradient (overlap) integral.}
Since the heavy axion's profile is thin compared with $R_s$, we set $r\simeq R_s$ throughout its support, and any slowly varying function of
$\chi$ can be evaluated at the
surface value $\chi_s\equiv\chi(R_s)$ up to fractional corrections
$\mathcal{O}(\ell_\mfa/R_s)$.  This gives
\begin{align}
\int_{R_{\rol}}^{R_s}\! r^2 W W_{,\chi}\,\mfa'^2\,dr
&\;\simeq\;
R_s^2\left(\frac{\Delta\mfa}{\ell_{\mfa}}\right)^{\!2}
\int^{R_s}_{\alpha R_{s}}\!dr\, W W_{,\chi}(\chi(r))
\nn\\
&\;\simeq\;
R_s^2\left(\frac{\Delta\mfa}{\ell_{\mfa}}\right)^{\!2}
\,W W_{,\chi}(\chi_s)\,\ell_\mfa.
\label{eq:grad-int-global}
\end{align}
The sign of $W W_{,\chi}(\chi_s)\propto \zeta$ therefore directly controls whether the
overlap term behaves as an effective stiffness ($W_{,\chi}(\chi_s)<0$) or an assisting drive
($W_{,\chi}(\chi_s)>0$).

\paragraph{Potential--slope integral.}
We decompose the potential slope into the usual single--field driving piece and
a genuine multi--field correction,
\begin{equation}
V_{,\chi}(\chi,\mfa)
   = V_{,\chi}(\chi_c,\mfa_-)
   + \big[V_{,\chi}(\chi,\mfa)-V_{,\chi}(\chi_c,\mfa_-)\big].
\end{equation}
The first term is the standard single--field contribution that drives the
dilaton across the shell of width $\Delta R$, while the second measures the
departure from the single--field case and is non-negligible only within the overlap
with the axion gradient, where both fields depart from their interior values.

Approximating $r\simeq R_s$ across both supports and evaluating slowly varying
functions at the surface values $\chi_s\equiv\chi(R_s^-)$ and
$\mfa_+$ gives
\begin{align}
\int_{R_{\rol}}^{R_s}\! r^2 V_{,\chi}&(\chi,\mfa)\,dr
\simeq
R_s^2\,V_{,\chi}(\chi_c,\mfa_-)\,\Delta R
\nn\\[4pt]
&\qquad+
R_s^2 \!\int^{R_s}_{\alpha R_s}\!dr\,
   \big[V_{,\chi}(\chi,\mfa)-V_{,\chi}(\chi_c,\mfa_-)\big]
\nn\\[4pt]
&\simeq
R_s^2\Bigl(V_{,\chi}(\chi_c,\mfa_-)\,\Delta R
+
\big[V_{,\chi}\big]^{\chi_s,\mfa_+}_{\chi_c,\mfa_-}\,
\ell_\mfa\Bigr).
\label{eq:Vchi_interaction}
\end{align}
This cleanly separates the usual single--field driving term from the gradient overlap
correction induced by the axion ramp.

\paragraph{Matter--coupling integral.}
Across the thin layer the density can be treated as constant,
$\rho_m(r)\simeq\rho_c$. This gives
\begin{align}
\int_{R_{\rol}}^{R_s}\!dr\,r^2\,A_{,\chi}(\chi)\rho_m(r)
\simeq& R_s^2\,A_{,\chi}(\chi_s)\,\rho_c\,\Delta R
\nn\\&= 6\,A_{,\chi}(\chi_s)\MPL^2\Phi_N(R_s)\,R_s\,\frac{\Delta R}{R_s},
\label{eq:matter-interaction}
\end{align}
where we used $\rho_cR_s^2\simeq6\MPL^2\Phi_N(R_s)$ to express the result in
terms of the surface Newtonian potential of the dense object.

\medskip
We note for later that the equations, \cref{eq:grad-int-global,eq:Vchi_interaction} show that the multi-field modification is controlled by the axion's ramp profile ($\Delta\mfa/\ell_{\mfa}$), which differs from object to object, implying the axion modification to the dilaton dynamics has a
generic dependence on object size and type.  Because the axion sector is tied to the matter density and does not generically couple gravitationally (and so universally) like the dilaton, composition can enter indirectly by shifting $\Delta\mfa$ and the effective ramp localisation, thereby inducing object-dependent environmental dynamics even when $A(\chi)$ is universal.

\subsubsection*{Charge and Shell profile relationships}\label{sec:Charge and Shell profile relationships}

The shell width can then be obtained by using the explicit evaluations
\cref{eq:grad-int-global,eq:Vchi_interaction,eq:matter-interaction} in \cref{eq:shell-balance}, yielding

\begin{equation}
\frac{\Delta R}{R_s}
\simeq
\frac{
2\Delta\chi
-
R_s\ell_{\mfa}
\left[
\Delta V_{,\chi}
+
W(\chi_s)W_{,\chi}(\chi_s)
\Big(\frac{\Delta\mfa}{\ell_{\mfa}}\Big)^2
\right]}
{
6\beta \MPL\,\Phi_N(R_s)
+ R_s^2 V_{,\chi}(\chi_c,\mfa_-)} ,
\label{eq:caseB}
\end{equation}
where
\[
\Delta V_{,\chi}
=
V_{,\chi}(\chi_s,\mfa_+)
-
V_{,\chi}(\chi_c,\mfa_-).
\]  
\Cref{eq:caseB} has an intuitive physical explanation. When $\chi_c$ and $\chi_{\rm env}$ are well--defined minima of the density--dependent effective potential, the total excursion $\Delta\chi$ is fixed. The axion contribution then modifies only how sharply the dilaton interpolates between these values near the surface. If the axion--induced contribution points in the same direction as the dilaton roll,
\begin{equation}
\big(W W_{,\chi}\big)(\chi_s)\,\Delta\chi > 0,
\end{equation}
then the ramp acts as an assisting source and the shell becomes narrower, reducing $\Delta R/R_s$. If the sign is opposite, the axion contribution opposes the roll and the shell must broaden to connect $\chi_c$ to $\chi_{\rm env}$.
Thus in the minima--based pinned regime the axion renormalises the shell thickness by biasing the near--surface evolution through the two--derivative interaction $W W_{,\chi}(\mfa')^2$.

\begin{figure}
    \centering
    \includegraphics[width=\linewidth]{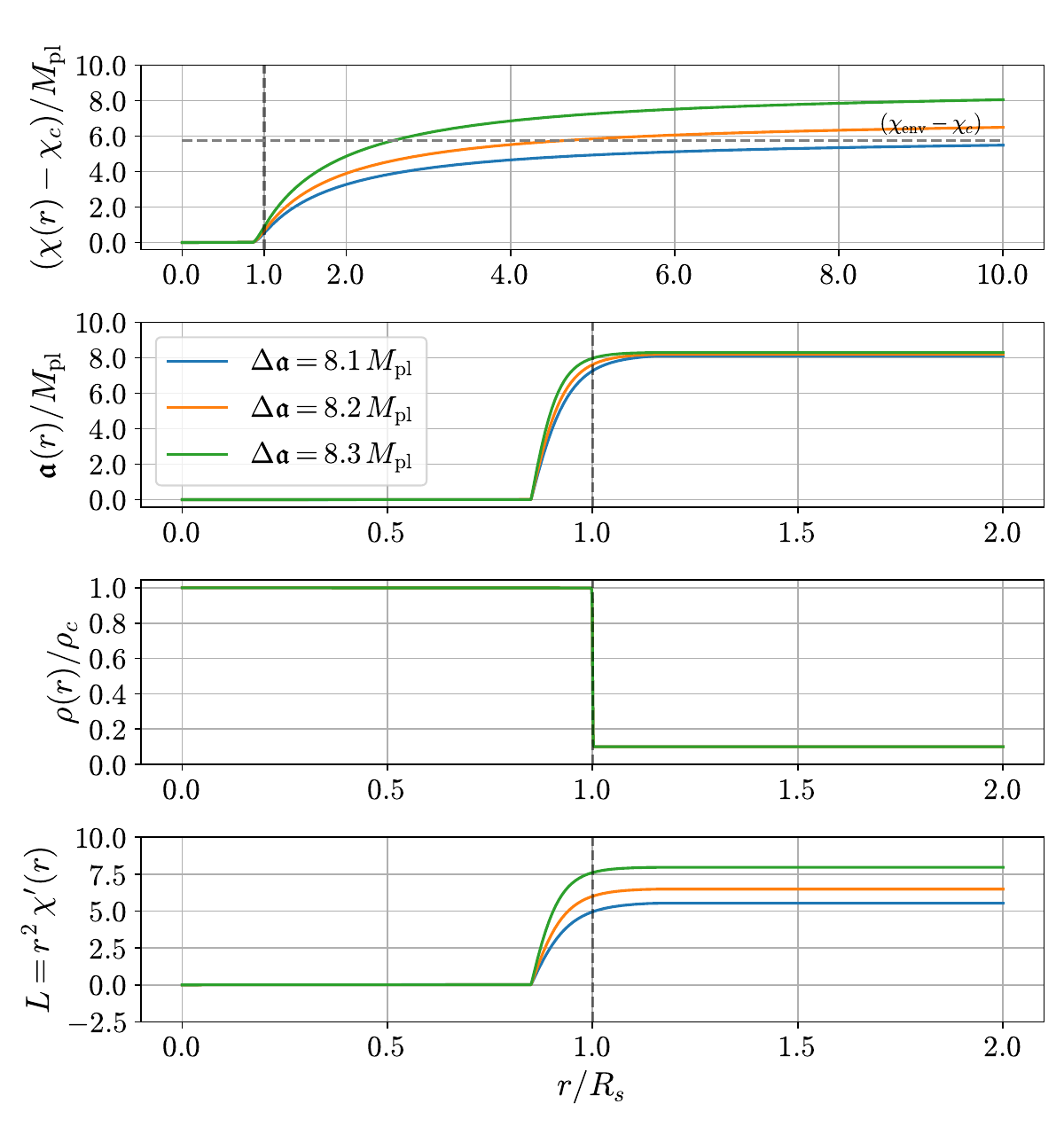}
    \caption{Radial profiles (\textbf{First:} dilaton, \textbf{Second:} axion, \textbf{Third:} density,
    \textbf{Fourth:} dilaton charge) for
    different axion boundary jumps $\Delta\mfa=\mfa_{+}-\mfa_{-}$ and a wide axion gradient using the flux conserving treatment for the axion described in \Appref{app:methodology} instead of solving the full equations of motion, with $\ell_\mfa = 0.15$. Here we took $\beta = 0.2$, $\zeta = \sqrt{2}$ and $W_0 = 1$. We also took boundary conditions for the dilaton field $\chi'(0) = 0$, $\chi(0) = \chi_c$ corresponding to the pinned regime. Crucially for numerical stability in the presence of exponential factors of the dilaton field, we choose an Earth--like
    density $\rho_c\simeq 10^{-9}\MPL^2/R_s^2$, with $V_0=0.5\,\rho_c$ and
    $\rho_{\rm env}=0.1\,\rho_c$ (an unrealistically small hierarchy).}
    \label{fig:thick_ramp_profiles}
\end{figure}

This effect is illustrated in \Cref{fig:thick_ramp_profiles}, where one can see the axion gradient driving the dilaton further and further towards its asymptotic value $\chi_{\rm env}$. At some point the axion can cause the dilaton to overshoot the asymptotic value, leaving a large residual offset between the field and $\chi_{\rm env}$ at large radius.

Now we turn to see how this relates to screening. Assuming a thin shell using \cref{eq:L_thin_shell} and that the dilaton approaches the Laplace tail exterior to the source as in \cref{eq:light_dilaton_yukawa_tail}, one has 
\begin{equation}
    \frac{R_s}{\Delta R}\left(\chi_s - \chi_c\right) = \chi_{\rm env} - \chi_s\,,
\end{equation}
and hence the exterior charge can be evaluated to
\begin{equation}\label{eq:thin_shell_charge}
    L = R_s\left(\chi_s -\chi_{\rm env}\right) =R_s \frac{\chi_c - \chi_{\rm env}}{1+\frac{\Delta R}{R}},
\end{equation}
which is independent of any explicit axion modification,
and given that $\chi_c$ and $\chi_{\rm env}$ are fixed minima of \cref{eq:single_field_eff_pot} this means that the external charge is only influenced by the axion through its effect on the shell width $\Delta R/R$, which is small by construction. This implies the axion sourcing has negligible effect on the external charge even if it drives the dilaton into the thin-shell configuration that the dilaton failed at doing on its own. This is because, while the axion is qualitatively pushing the dilaton into a more charge-suppressed configuration for the dilaton in isolation, this external axion gradient comes with its own charge sourcing on the dilaton through the coupling term in \cref{eq:shell-balance}, removing the ability for the net charge to be suppressed. 

\begin{figure}
    \centering
    \includegraphics[width=\linewidth]{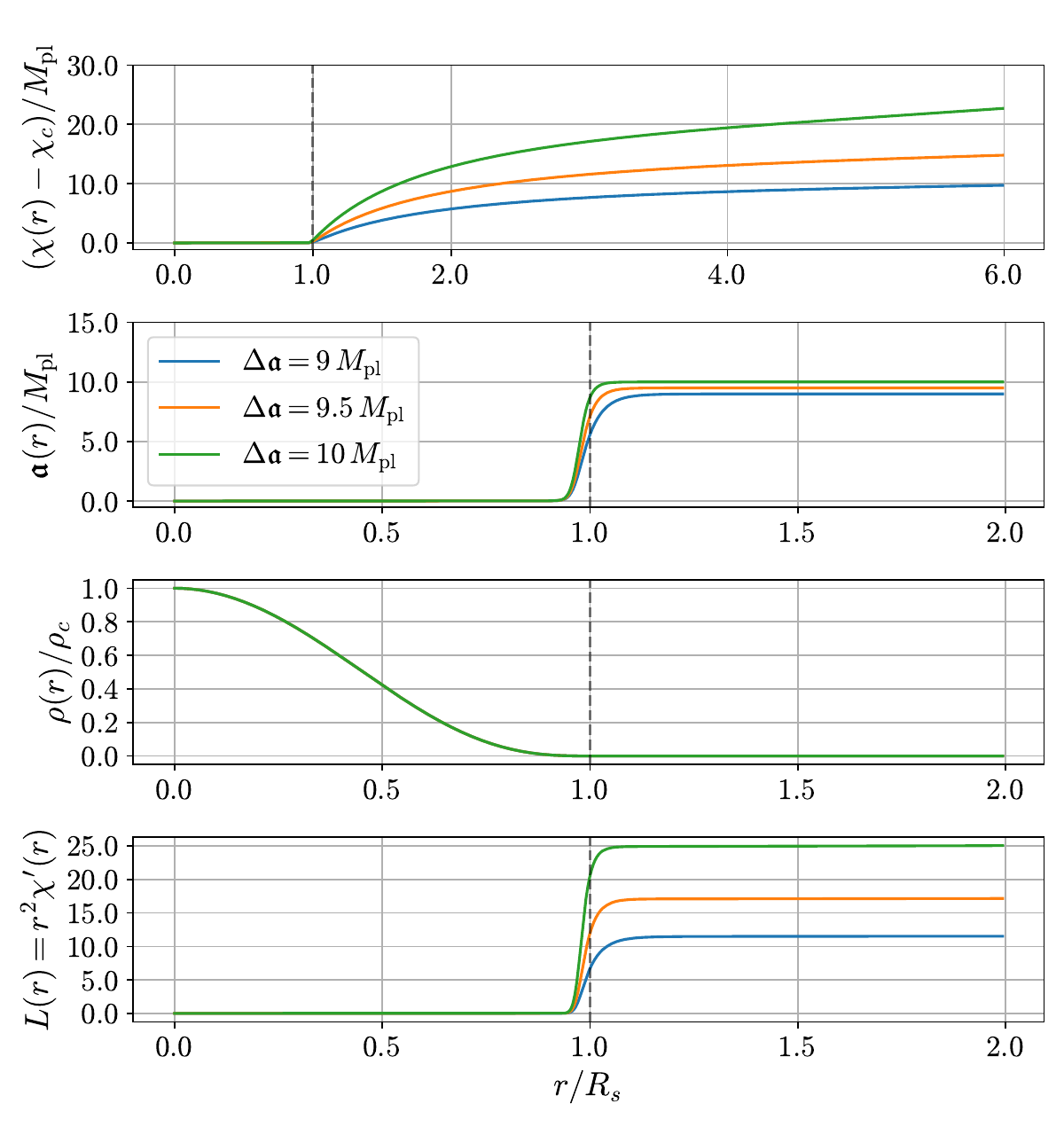}
    \caption{Radial profiles for the same parameters as in \Cref{fig:thick_ramp_profiles} except for $\rho_{\rm env} = 10^{-8} \rho_c$ and a star-like object density profile described in \Appref{app:methodology}. In this case we solve the full multi-field equations using the microphysical parameters $\Lambda_\mfa = 3\times10^{-6}\MPL$ and $m_\mfa = 3\times10^{-16}\,\text{eV}$.}
    \label{fig:star_thick_ramp_profiles}
\end{figure}

This is numerically confirmed in \Cref{fig:thick_ramp_profiles} as well for wider axion ramps than the analytic approximations accommodate (showing the results do not change if this assumption is relaxed), where we can see in the bottom panel that driving the dilaton to its asymptotic value only changes the scalar charge by $\mathcal{O}(1)$ factors. It in fact increases the scalar charge as the shell width is made more thin, as predicted from \cref{eq:thin_shell_charge}. The same effect is shown to occur for star-like density profiles in \Cref{fig:star_thick_ramp_profiles}, showing that this result is robust across different density distribution and axion ramp effects.
Sourcing the dilaton to make its transition close to the surface therefore allows for a thinner shell, but does not assist in any way with reducing its exterior charge. 

\subsection{Thin shell for ultra-light, unpinned dilatons}
\label{sec:thin_shell_unpinned}

The thin--shell mechanism relies on the existence of two density--dependent
equilibria of the dilaton effective potential, an interior value $\chi_c$ and an
exterior value $\chi_{\rm env}$, between which the field interpolates within a
parametrically thin surface layer.  This picture breaks down when the dilaton is
ultra--light on astrophysical scales.

For $m_\chi R_s \ll 1$, the dilaton cannot relax to a dense--matter equilibrium
inside a finite object. Instead it stays close to the environmental value
$\chi_{\rm env}$ throughout the interior, with only $\mathcal{O}(\Phi_N)$
deviations, so there is no interior plateau and no parametrically thin rolling
layer even in the single--field theory. The same effective behaviour is
expected for yoga dilatons introduced in \Cref{sec:axio-dilaton_setup}, because although their potential has a shallow
dark--energy minimum that fixes the asymptotic value cosmologically, its
curvature remains far too small to pin the field in astrophysical environments,
so one effectively has $m_{\rm eff}R_s\ll 1$ and the thin--shell picture is not
well defined.

\begin{figure}
    \centering
    \includegraphics[width=\linewidth]{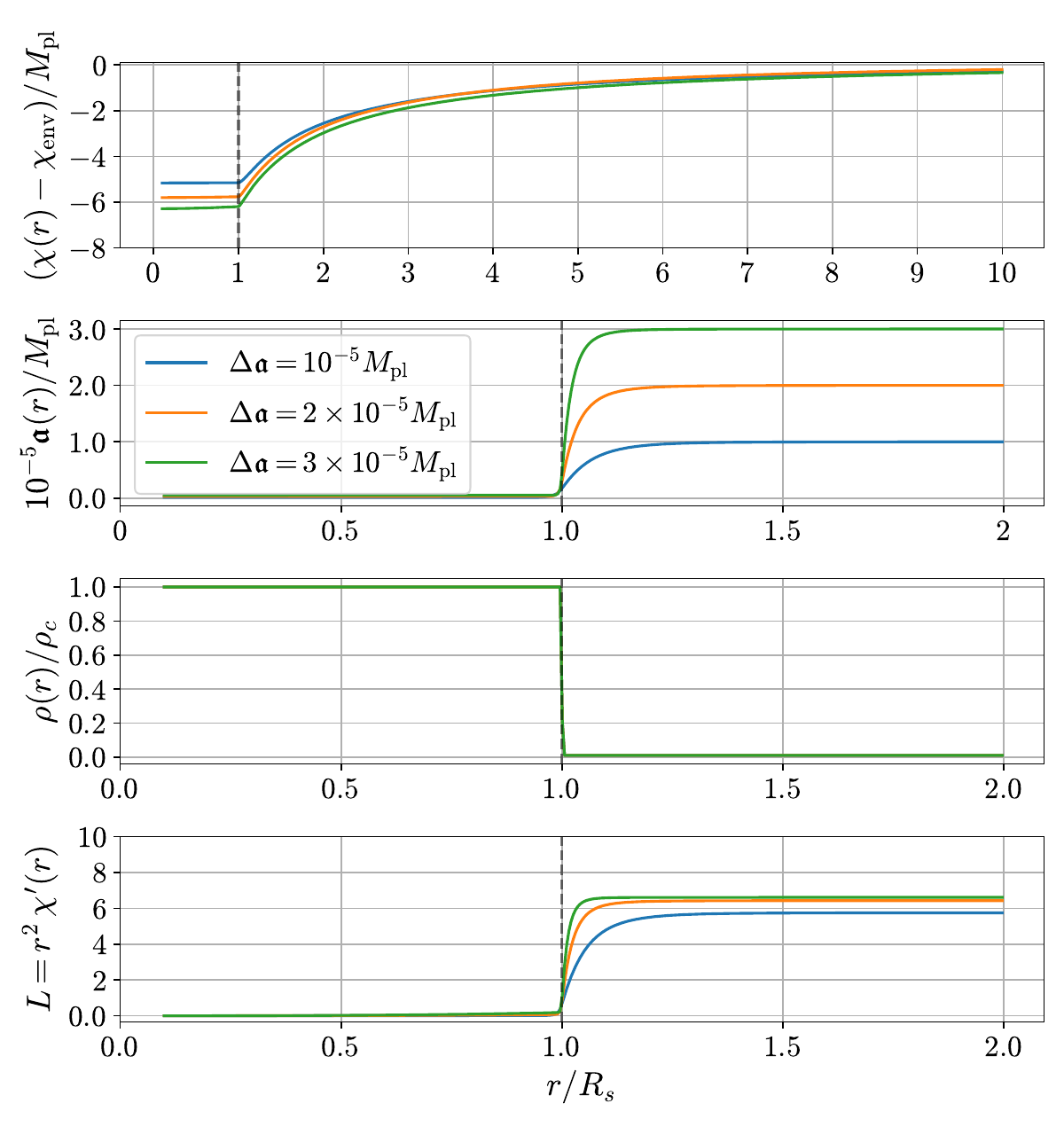}
    \caption{Radial profiles for an unpinned dilaton with axion
    boundary jumps $\Delta\mfa=\mfa_+-\mfa_-$. We used the same parameters as \Cref{fig:thick_ramp_profiles} but swapped a boundary condition at $\chi(0) = \chi_c$ with an asymptotic boundary condition $\chi(r_{\rm max}) = \chi_{\rm env}$. No parametrically thin rolling layer
    forms, consistent with $m_\chi R_s\ll1$.  Axion gradients modify the global
    profile rather than suppressing the exterior scalar charge.}
    \label{fig:profiles_stacked_small_mass_thin_shell}
\end{figure}

This behaviour is illustrated in
\Cref{fig:profiles_stacked_small_mass_thin_shell}.  In this unpinned regime,
localised axion gradients do not compress the radial width of the dilaton's transition. Rather, they modify the global dilaton profile, generically increasing
the total field excursion and enhancing the exterior scalar charge $L$ relative
to the single--field case, as described by \cref{eq:light_dilaton_yukawa_tail}.

Thin--shell screening therefore fails for ultra--light dilatons even in the
presence of axion--induced near--surface dynamics.  Screening in this regime must
instead suppress the net scalar charge directly, rather than relying on geometric
localisation of the profile.

In the next section we show that this is achieved by the BBQ mechanism, in which
the axion backreaction dynamically selects a configuration that minimises the
total energy and reduces the exterior charge.

\section{BBQ screening:  gradient suppression when the dilaton rolls freely (and $W_{,\chi}(\chi_s)<0$)}
\label{sec:axio-dilaton_no_minimum}

Having established that thin-shell screening of a light dilaton from solar system tests of gravity is seemingly not viable, regardless of pinning complexities in the multi-field context, we now focus on the opposite direction one can take.  The next question to ask is what if instead the axion does something other than sourcing an additional dilaton profile? What happens if the axion is directly \emph{suppressing} the dilaton's gradient, further frustrating it from reaching its environmental minimum exterior to objects.

\subsection{Pinned gradient suppression}
\label{sec:axion_gradient_suppression}

\begin{figure}
    \centering
    \includegraphics[width=\linewidth]{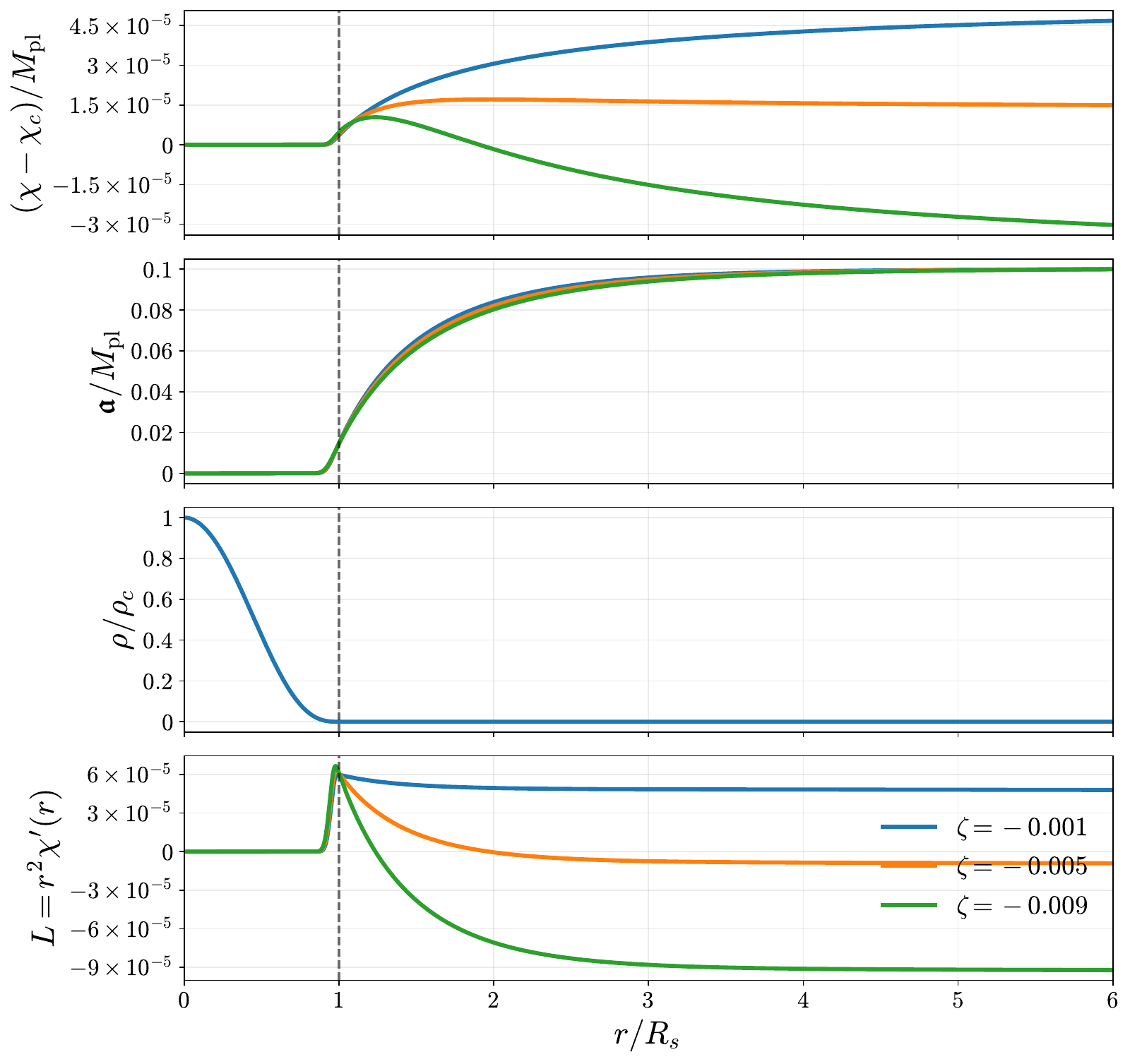}
    \caption{Field profiles and dilaton charge evaluated for a pinned scalar in the presence of a suppressing coupling to the axion in a stellar density profile outlined in \Appref{app:methodology}. Here we took $W_0 = 1$, $V_0 = 10^{-5}\rho_c$ and $\rho_{\rm env} = 10^{-15}\rho_c$. We solve the full coupled two-field system using the microphysical axion potentials \cref{eq:U_ax} and \cref{eq:V_ax} with $\Lambda_\mfa = 3\times10^{-7}\MPL$ and $m_\mfa = 3\times10^{-16}\,\text{eV}$.}
    \label{fig:star_pinned_dilaton_gradient_suppression}
\end{figure}

\Cref{fig:star_pinned_dilaton_gradient_suppression} illustrates the behaviour of the massive,  pinned-dilaton system when the axion--dilaton kinetic coupling satisfies $W_{,\chi}\propto\zeta<0$, inducing a negative sourcing contribution the dilaton's Klein-Gordon \cref{eq:eom_dilaton}. In this case the axion gradient does not assist the dilaton in rolling towards its environmental minimum $\chi_{\rm env}$, but instead acts as an effective stiffness that opposes its motion. The resulting backreaction pushes the dilaton back up its exponential potential and flattens the exterior profile relative to the single--field expectation. 

As shown in the bottom row of \Cref{fig:star_pinned_dilaton_gradient_suppression}, it is possible to tune the axion coupling such that this effect partially cancels the dilaton scalar charge, thereby reducing the associated fifth force. However, the figure already makes clear that this cancellation is highly delicate. A slightly stronger coupling reverses the effect and instead enhances $|L|$, as the axion begins to drive the dilaton against its intrinsic rolling direction.

This behaviour can be understood directly from the global expression for the dilaton charge \cref{eq:global-L-def},
which shows that the exterior charge receives contributions from the potential slope, the matter coupling, and the axion--induced kinetic term. For $\zeta<0$ the kinetic contribution can oppose the flux generated by the potential and matter terms near the surface. Achieving compatibility with solar--system bounds therefore requires these contributions to cancel at the level of at least three orders of magnitude, so that 
\begin{equation}
    \frac{L}{L_0} = \frac{\beta_{eff}}{\beta}\sim10^{-3}.
\end{equation}

The reason this cancellation is so finely tuned in the present setup is that the dilaton is massive and remains pinned close to the minimum of its effective potential throughout the interior of the object. Although dilaton gradients are energetically costly, displacing a heavy field away from this minimum incurs a large potential--energy penalty over the entire volume. The equilibrium configuration therefore necessarily contains an intrinsic surface gradient fixed by the mismatch between the interior and exterior minima. Any additional interaction, such as the axion kinetic coupling, must then be tuned independently to cancel this pre--existing flux across the configuration rather than being selected dynamically by energy minimisation.

In other words, the axion gradient is being used to counteract a surface gradient that is itself enforced by pinning. The cancellation required to suppress the charge is therefore accidental rather than structural, leaving little parametric room for successful screening. The very mechanism that renders the dilaton heavy and pinned is therefore also what obstructs robust gradient suppression.

As for how this relates to the discussion above and the apparent inconsistency that the axion can affect the exterior charge through \cref{eq:global-L-def} while \cref{eq:thin_shell_charge} suggests that $L$ is essentially axion--independent, the key point is that the thin--shell derivation implicitly assumes that the dilaton remains close to the minimum of its density--dependent effective potential throughout the interior and up to radii very near the surface. In the pinned thin-shell regime the intrinsic surface gradient is fixed by the mismatch between the interior and environmental minima, and the axion enters the shell analysis only as a perturbative modification of the near--surface dynamics, effectively renormalising the shell width $\Delta R/R_s$ while leaving the exterior charge controlled primarily by the fixed excursion $\chi_c-\chi_{\rm env}$. By contrast, the regime in which $W_{,\chi}<0$ suppresses the charge requires the axion contribution in \cref{eq:global-L-def} to be comparable to the intrinsic flux generated by the potential and matter terms. Once this occurs the axion backreaction necessarily displaces the dilaton away from the effective minimum in the near--surface region, so the field no longer tracks the pinned solution assumed in deriving \cref{eq:thin_shell_charge}. The analytic thin--shell/minima expansion therefore ceases to be self--consistent in this regime. Charge suppression is then possible only through a cancellation among the different contributions to the global charge relation, which is inherently delicate and sensitive to the precise value of the axion coupling.

\subsubsection*{Environmentally-dependent screening}

\begin{figure}
    \centering
    \includegraphics[width=\linewidth]{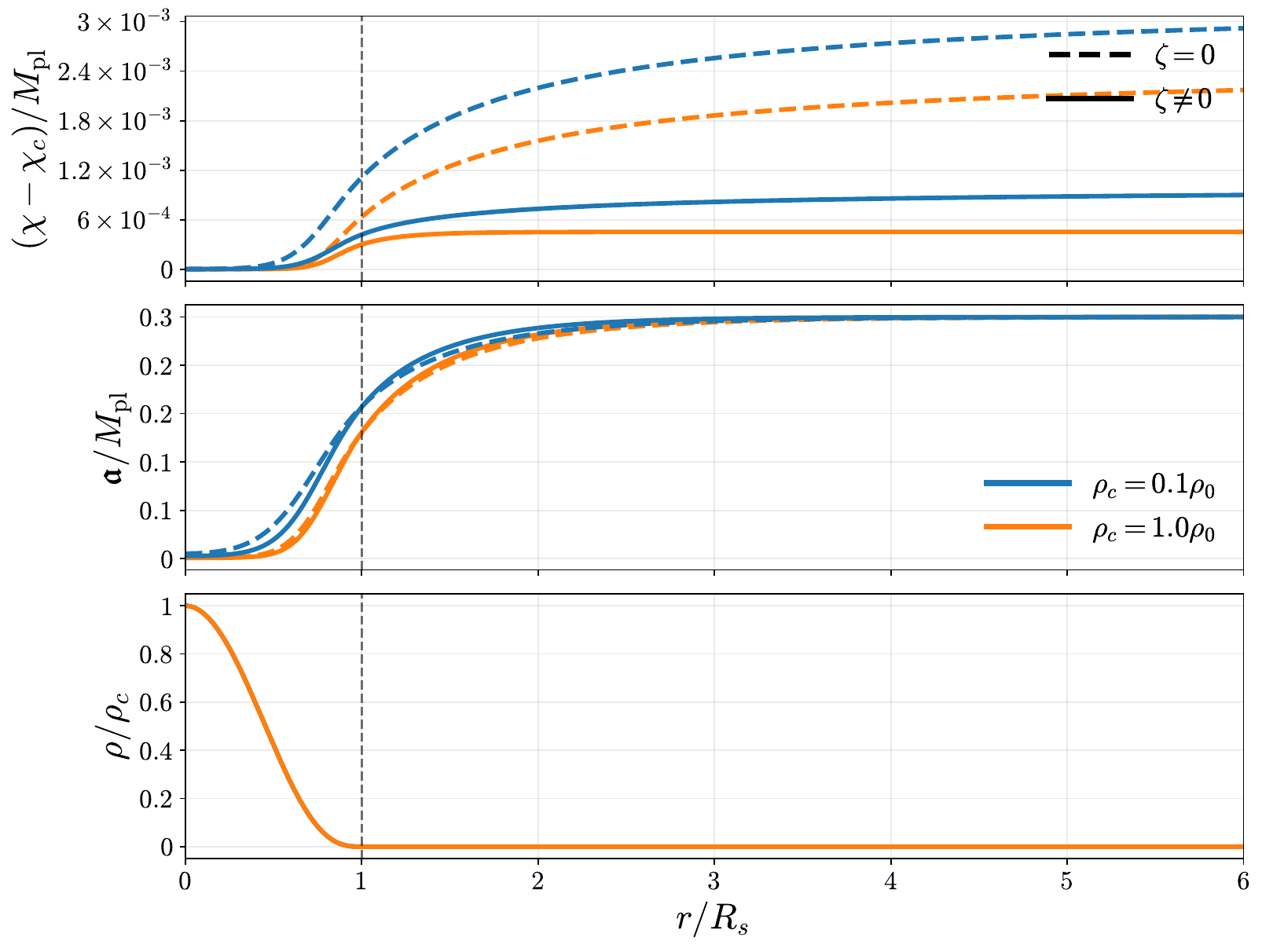}
    \caption{Profiles of the constituent fields of the pinned dilaton system when the axion-dilaton coupling $\zeta$ has been tuned to cancel the dilaton gradient enough to satisfy solar system tests for a density distribution of the Sun in blue. The lines in orange show the profiles for exactly the same density distribution but $10\%$ of the core density. Dashed lines show the corresponding $\zeta = 0$ cases for each. $\rho_c$ is the core solar density and here we took $V_0 = 10^{-3}\rho_c$, $\rho_{\rm env} = 10^{-15}\rho_c$, and $W_0 = 1$ and the two-field system was solved exactly with axion microphysical parameters $\Lambda_\mfa = 3\times10^{-6}\MPL$ and $m_\mfa = 3\times10^{-16} \,\text{eV}$.}
    \label{fig:environmental_profiles}
\end{figure}

\begin{figure}
    \centering
    \includegraphics[width=\linewidth]{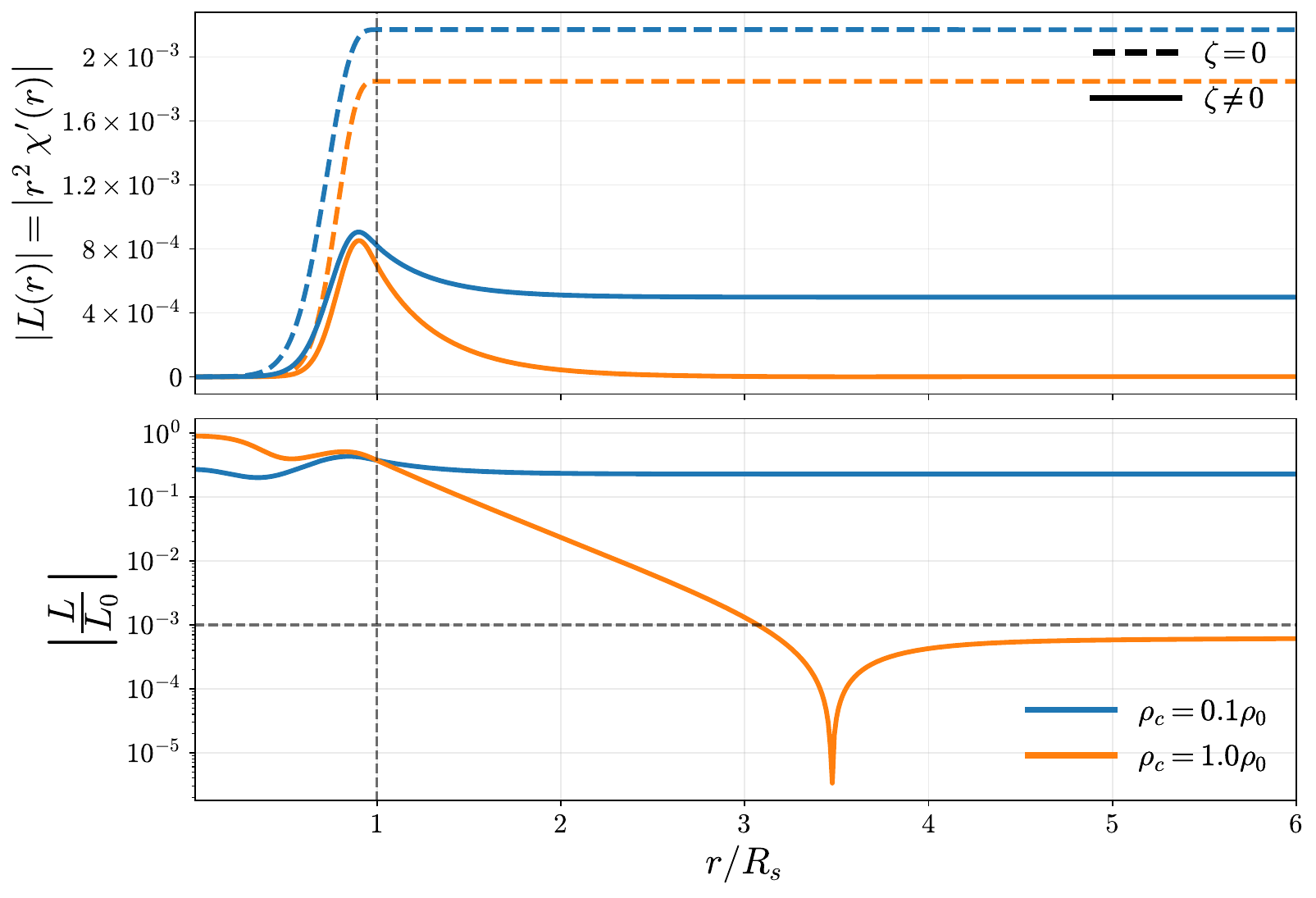}
    \caption{\textbf{Top:} dilaton charge for the corresponding relevant profiles in \Cref{fig:environmental_profiles}. \textbf{Bottom:} Charge relative to the unscreened charge $L_0$ obtained by running the simulation with $\zeta = 0$.}
    \label{fig:environmental_L}
\end{figure}

This tuning problem is exacerbated once one demands screening across multiple objects.
An axion gradient tuned to cancel the dilaton's surface-generated flux for a given density profile
will not generically cancel it with the same efficiency in another object with different $\rho_m(r)$,
because the relative weighting of the terms in \cref{eq:global-L-def} changes with the interior profile.
This is illustrated in \Cref{fig:environmental_profiles,fig:environmental_L}, which choose the axion coupling $\zeta$
so that the exterior charge is strongly reduced ($L/L_{0} \sim 10^{-3}$ far outside the source) for a solar-density source in blue, and then show this does not produce a
comparable suppression for an otherwise identical star with $\rho_c = 0.1\,\rho_\odot$ in orange, which retains
a much larger residual charge.

In a solar-system setting this lack of universality is fatal: satisfying solar constraints that probe the
Sun's effective coupling (e.g.\ Shapiro time-delay bounds during solar conjunction) does therefore not ensure that
other bodies are also screened. For example, if the Earth and Moon retain appreciable scalar charges, differential
accelerations and equivalence-principle violations would be tightly constrained by lunar-laser-ranging \cite{Williams:2004qba, Williams:2012nc}
and planetary-ephemeris tests \cite{Fienga:2009ub,Fienga:2023ocw}, regardless of whether the Sun itself has been tuned to be screened.

For this reason, axion--induced suppression of the dilaton charge is generically unnatural in the pinned regime. In the next section we instead consider light, unpinned dilatons, for which the surface value of the field is not fixed by a local minimum. In that case the axion backreaction can participate directly in the global energy minimisation of the configuration (because it is not so costly to displace the dilaton's interior value any more, energetically speaking), allowing charge suppression to arise dynamically rather than through fine-tuning object-by-object.

\subsection{Un-pinned gradient suppression}

The preceding section shows promise, and there is only one assumption being made that is neither realistic for a dark-energy-scale dilaton, nor helpful for screening: that the dilaton is pinned on astrophysical scales. Relaxing this assumption changes the screening picture entirely.

Doing so requires moving away from extremising the effective potential \cref{eq:full_exponential_potential} to define an equilibrium configuration, and instead allowing the surface and asymptotic values to float. As emphasised in Ref.~\cite{Brax:2010gi}, screening can still occur when no such equilibria exist, and the relevant configuration is selected by minimising the total static energy subject to continuity of the field and its radial flux. This is precisely the mechanism already identified to work for yoga dilatons, and here we generalise the narrow-width (infinitely sharp) axion-gradient treatment to finite-width (but still thin compared to the dilaton) linear ramps and fully coupled numerical solutions to compare with the regimes above.

\subsection{Global scalar charge}

Outside the source the dilaton satisfies the Laplace equation and takes the
form \cref{eq:light_dilaton_yukawa_tail},
where the scalar charge is obtained from the radial flux as in \cref{eq:global-L-def}.
Assuming the axion gradient is approximately localised to near the surface and the dilaton does not deviate from its surface value within the body appreciably yields
\begin{align}
  L \simeq\;
    \frac{R_{\rm roll}^3}{3}V_{,\chi}(\chi_c,\mfa_-)
  +& 2\beta\MPL\,\Phi_N(R_s)\,R_s
  \nn\\&+ R_s^2\Big(\frac{\Delta\mfa}{\ell_{\mfa}}\Big)^2
    W(\chi_s)W_{,\chi}(\chi_s)\,\ell_\mfa,
  \label{eq:L-three-terms}
\end{align}
where the second term defines the single–field charge as in \cref{eq:betaeff_def_general} and the final term is the axion–induced
surface contribution arising purely from the two–derivative
$\sigma$–model structure. The scalar charge depends on the boundary value $\chi_s$ and the central value $\chi_c$. They are not independent variables. Indeed, this emanates from the way the solutions to the dynamical equations are obtained, i.e.  by specifying the value of $\chi_c$ and imposing that $\chi'=0$ at the origin. This implies the dependence $\chi_s(\chi_c)$. Varying $\chi_s$ therefore provides an accurate
proxy for varying $\chi_c$ when selecting the equilibrium configuration.

\subsection{Energy decomposition}

The total static energy for a spherically symmetric configuration is
\begin{align}\label{eq:e_tot}
  E &= 4\pi\!\int_0^\infty dr r^2\Big[
     \tfrac12\chi'^2
     + \tfrac12 W^2\mfa'^2
     + V(\chi,\mfa)
     + A(\chi)\rho_m(r)
     \Big]
     \nn\\&= E_{\rm in} + E_{\rm sh} + E_{\rm out},
\end{align}
where we have split the energy up by the regions from which it originates, the interior ($E_{\rm in}$), the axion shell ($E_{\rm sh}$) and the exterior ($E_{\rm out}$). In the interior region the fields do not vary much and this gives
\begin{equation}\label{eq:E_in}
  E_{\rm in}
  = \frac{4\pi}{3}R_s^3\Big[V(\chi_c,\mfa_-)
     + A(\chi_c)\rho_c\Big],
\end{equation}
which is independent of $\chi_s$.

In the shell region $\alpha R_s<r<R_s$ the axion dominates the energy budget due to its fast transition near the surface.  The shell
energy is
\begin{align}
  E_{\rm sh}
  = 4\pi\!\int_{\alpha R_s}^{R_s}\!dr\, r^2
    \Big[
      \tfrac12(\chi')^2
      + \tfrac12 W^2(\chi)(\mfa')^2
      &+ V(\chi,\mfa)
      \nn\\&+ A(\chi)\rho_c
    \Big].
\end{align}
If the axion dominates the varying energy density in the layer and $W(\chi)$
varies slowly across the shell, then to leading order
\begin{equation}
  E_{\rm sh}(\chi_s)
  \simeq
  2\pi R_s^2\,W^2(\chi_s)
  \frac{(\Delta\mfa)^2}{\ell_{\mfa}^2}\,\ell_{\mfa}
  + E_0,
  \label{eq:EShell-approx}
\end{equation}
where $E_0$ contains the dilaton contribution independent of $\chi_s$ and dependent on the central value $\chi_c$.

\begin{figure}
    \centering
    \includegraphics[width=\linewidth]{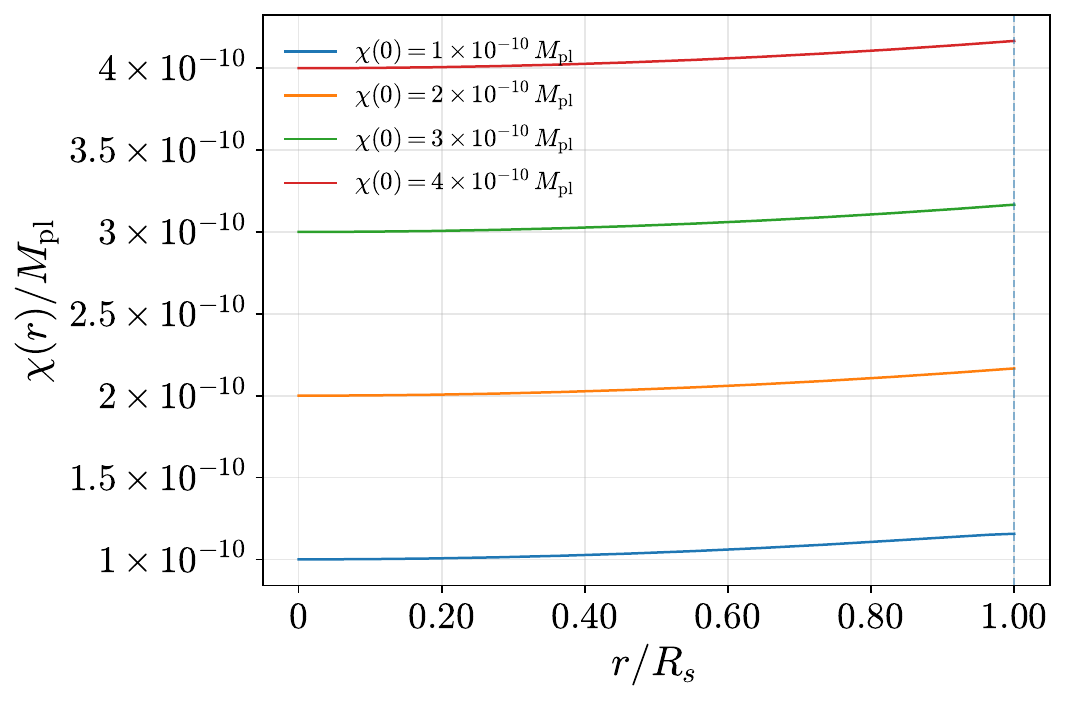}
    \caption{Radial evolution of the dilaton field interior to an idealised constant density object (in the absence of an axion) for different central dilaton values $\chi_c$, showing the almost exact correspondence between the surface and central values. Here we took $\rho_c$ to correspond to the surface density of the Earth with $V_0 = 0.5 \rho_c$ and $\rho_{\rm env} = 0.1 \rho_c$ and $W_0 = 1$.}
    \label{fig:chi_trajectories}
\end{figure}

In the exterior, where $\chi' = L/r^2$, the gradient energy is
\begin{equation}
  E_{\rm out}^{(\nabla\chi)}
  = 2\pi\,\frac{L^2}{R_s},
\end{equation}
and the potential term contributes only a constant shift. 
The explicit
$\chi_s$–dependent part of the total energy is therefore
\begin{equation}
  E_{\rm tot}(\chi_s)
  = 2\pi R_s^2\,W^2(\chi_s)
    \frac{(\Delta\mfa)^2}{\ell_{\mfa}^2}\,\ell_\mfa
  + 2\pi\,\frac{L^2}{R_s}.
  \label{eq:Etot-compact}
\end{equation}
Deriving \cref{eq:Etot-compact} requires neglecting the explicit interior
contribution \cref{eq:E_in}, which depends primarily on the central value
$\chi_c$. As stated previously, however, the surface and central values are tightly
correlated, as illustrated in
\Cref{fig:chi_trajectories}. In principle these terms should be retained for complete precision; however, extremising the energy would then require the explicit functional forms of $V(\chi,\mfa)$ and $A(\chi)$, which this derivation is agnostic about. We therefore drop these terms and verify their negligibility in the regimes of interest below.

\subsection{Energy minimisation and suppression of the dilaton charge}
\label{sec:bbq_energy_min}

The energy depends on a single variable , i.e. the central value $\chi_c$, as the boundary value $\chi_s$ is a function of $\chi_c$ obtained by solving the dynamical equations of motion. In practice screening will impose that the terms in the axion contribution proportional to $\Big(\frac{\Delta\mfa}{\ell_{\mfa}}\Big)^2$ are large implying that one can neglect the variation of the terms involving the interior value $\chi_c$ of the dilaton and minimise directly by varying the surface value $\chi_s$. 
We now derive this surface value $\chi_s$ by minimising the total energy
\cref{eq:Etot-compact}.  To make the minimisation algebraic, and in line
with the examples used in \Cref{sec:thin-shell-piecewise-linear}, we take the exponential form for $W(\chi)$ \cref{eq:exp_W}
(which is also representative of any monotonic kinetic prefactor relevant for the
axion ramp). Appendix~\ref{app:quadraticW} illustrates the corresponding calculation for a
quadratic $W(\chi)$, showing that the structure of the result is generic.
The key point is that $L$ itself depends on
$\chi_s$ through the axion--induced surface contribution to the flux balance
\cref{eq:L-three-terms}, which in this regime is controlled by the
two--derivative interaction $W(\chi)W_{,\chi}(\chi)(\mfa')^2$.

Minimising \cref{eq:Etot-compact} with respect to $\chi_s$ yields the algebraic
condition
\begin{equation}
    \frac{R_s \ell_\mfa}{\ell_{\mfa}^2}\,
    \frac{(\Delta\mfa)^2}{\MPL^2}\,
    W^2(\chi_s)
    = -\,\frac{1}{2\zeta^2}\left(4\zeta\beta\,\Phi_N(R_s) + 1\right).
    \label{eq:minimisation-clean}
\end{equation}
Because the left-hand side is manifestly positive, \cref{eq:minimisation-clean}
has a solution only if
\begin{equation}
  4\zeta\beta\,\Phi_N(R_s)+1<0\,,
  \qquad \text{i.e.}\quad \zeta\beta<0\ \ \text{and}\ \ |4\zeta\beta\Phi_N|\gtrsim 1\,.
  \label{eq:bbq_sign_condition}
\end{equation}
Which is saying physically, $W(\chi)W_{,\chi}(\chi)$ must oppose the
matter--induced pull in the flux balance so that the system can lower its total
energy by reducing the exterior gradient energy $L^2/R_s$, and it must oppose the
matter--induced pull by a large amount, hence a strong coupling between the axion and dilaton is required.

For an exponential $W$,
\cref{eq:minimisation-clean} determines the surface value of the dilaton
$\chi_s$ explicitly:
\begin{equation}
    \chi_s
    = \frac{\MPL}{2\zeta}
      \ln\!\left[
      -\,\frac{\ell_{\mfa}}{R_s}
        \frac{\MPL^2}{(\Delta\mfa)^2}
        \frac{1}{2\zeta^2}
        \frac{4\zeta\beta\,\Phi_N(R_s) + 1}{W_0^2}
      \right],
    \label{eq:chis-BBQ-final}
\end{equation}
with the logarithm understood to be real precisely when the sign condition
\cref{eq:bbq_sign_condition} is satisfied.

\begin{figure*}
  \centering
  \subfloat{%
    \includegraphics[width=0.47\textwidth]{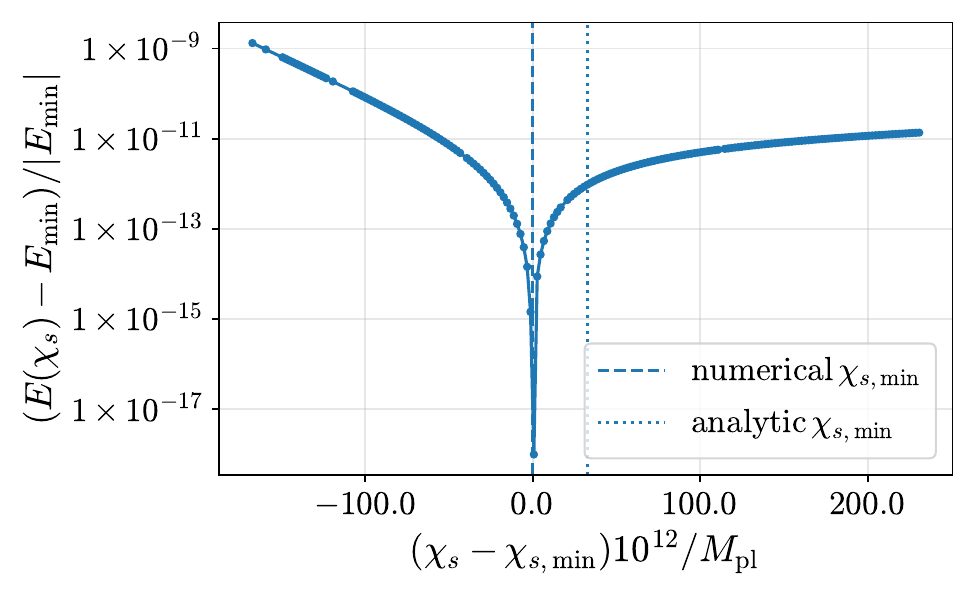}
  }
  \hfill
  \subfloat{%
    \includegraphics[width=0.47\textwidth]{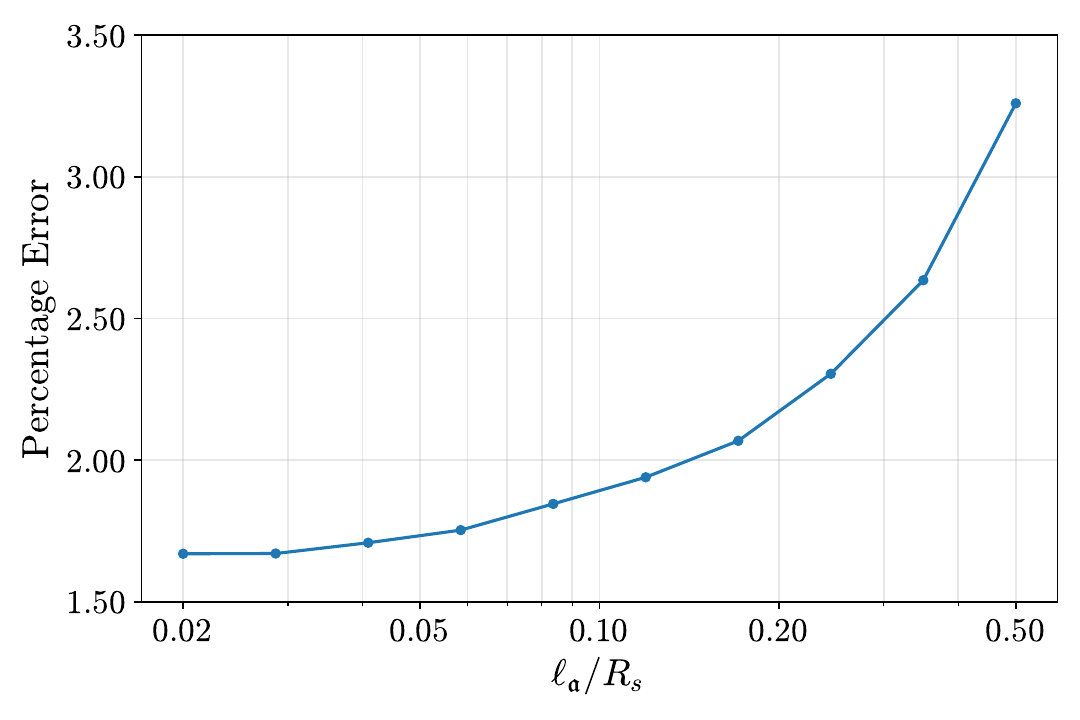}
  }
  \caption{\textbf{Left:} Relative difference between the Total system energy $E(\chi_s)$ and the minimum evaluated energy $E_{\rm min}$ as a function of $\chi_s$ for the green curve in \Cref{fig:profiles_stacked_bbq}, corresponding to a wide axion gradient with $\ell_\mfa = 0.4$. We also plot the analytic prediction for the energy minimising $\chi_s$ from \cref{eq:chis-BBQ-final} as a dotted line, compared to the numerically obtained minimum in the dashed line. \textbf{Right:} Percentage error in the analytic estimate \cref{eq:chis-BBQ-final} compared to the numerically obtained minimum energy $\chi_s$ for different axion ramp widths.}
  \label{fig:analytic_vs_numerical}
\end{figure*}

The derivation of \cref{eq:chis-BBQ-final} assumes that the dominant
$\chi_s$ dependence of the total energy arises from the axion-driven
surface contribution, while the implicit dependence entering through the
interior value $\chi_c$ is subleading. The validity of this approximation can be tested directly by performing a
full numerical energy minimisation of the coupled system, the methodology of which is discussed in \Appref{app:methodology}. This is shown in
\Cref{fig:analytic_vs_numerical}, where the total energy
\cref{eq:e_tot} is evaluated numerically as a function of $\chi_s$.
The analytic prediction \cref{eq:chis-BBQ-final} reproduces the location of
the numerical minimum to within $\lesssim 10^{-11}\MPL$ across the range of
axion ramp widths considered. We can see that the approximation is best for thinner axion gradients and becomes inaccurate as one considers more gradual axion ramps, going beyond the fast transition window approximation made in the analytic derivation. This is because the dominant energy contribution becomes no-longer localised within a small shell at the surface, as was assumed in \cref{eq:EShell-approx}, and one needs to include additional corrections from contributions outside this shell width as the axion gradient widens.
We have further verified that this residual
offset produces no observable change in the inferred charge suppression, and
the resulting $L/L_0$ (and hence $\beta_{\rm eff}$) remains unchanged at the
level relevant for the screening conclusions of this work.

Inserting \cref{eq:minimisation-clean} into the flux expression
\cref{eq:L-three-terms} (and using $W_{,\chi}=(\zeta/\MPL)W$ for the
exponential prefactor) gives
\begin{equation}
  \frac{L}{L_0}
  =
  \frac{
      -\tfrac{\MPL}{2\zeta}
      + \tfrac{R_{\rm roll}^3}{3R_s}\,V_{,\chi}(\chi_{\rm c},\mfa_-)
      }
      {2\beta \MPL\,\Phi_N(R_s)}\,,
  \label{eq:L-L0-final-clean}
\end{equation}
showing explicitly how the axion--induced surface term suppresses the usual
single--field contribution to the charge, up to a remainder controlled by
$1/\zeta$ and by the (model-dependent) potential contribution from $V_{,\chi}$.
This shows that screening can occur even when there is no parametrically
thin rolling layer: what is reduced is the \emph{amplitude} of the exterior
$1/r$ tail, not necessarily the radial width over which $\chi$ varies.

Successful screening then requires an extremely large \emph{negative} value for $\zeta$ given that $V,_\chi$ should independently be small for a cosmologically light dilaton, proving that screening works if $\frac{W,_\chi}{W} < 0$. If $\Phi_N \sim 10^{-6}$, corresponding to the Solar surface Newtonian potential, and imposing a reduction of the effective coupling by a factor of the order of $10^3$ to pass solar system tests of gravity, one then requires $\left|\zeta\right| \gtrsim 10^9$ for such suppression of the exterior charge for $\beta\sim \mathcal{O}(1)$. We will return to all of the problems associated with realising this in a successful model in the next section.

\begin{figure}
    \centering
    \includegraphics[width=\linewidth]{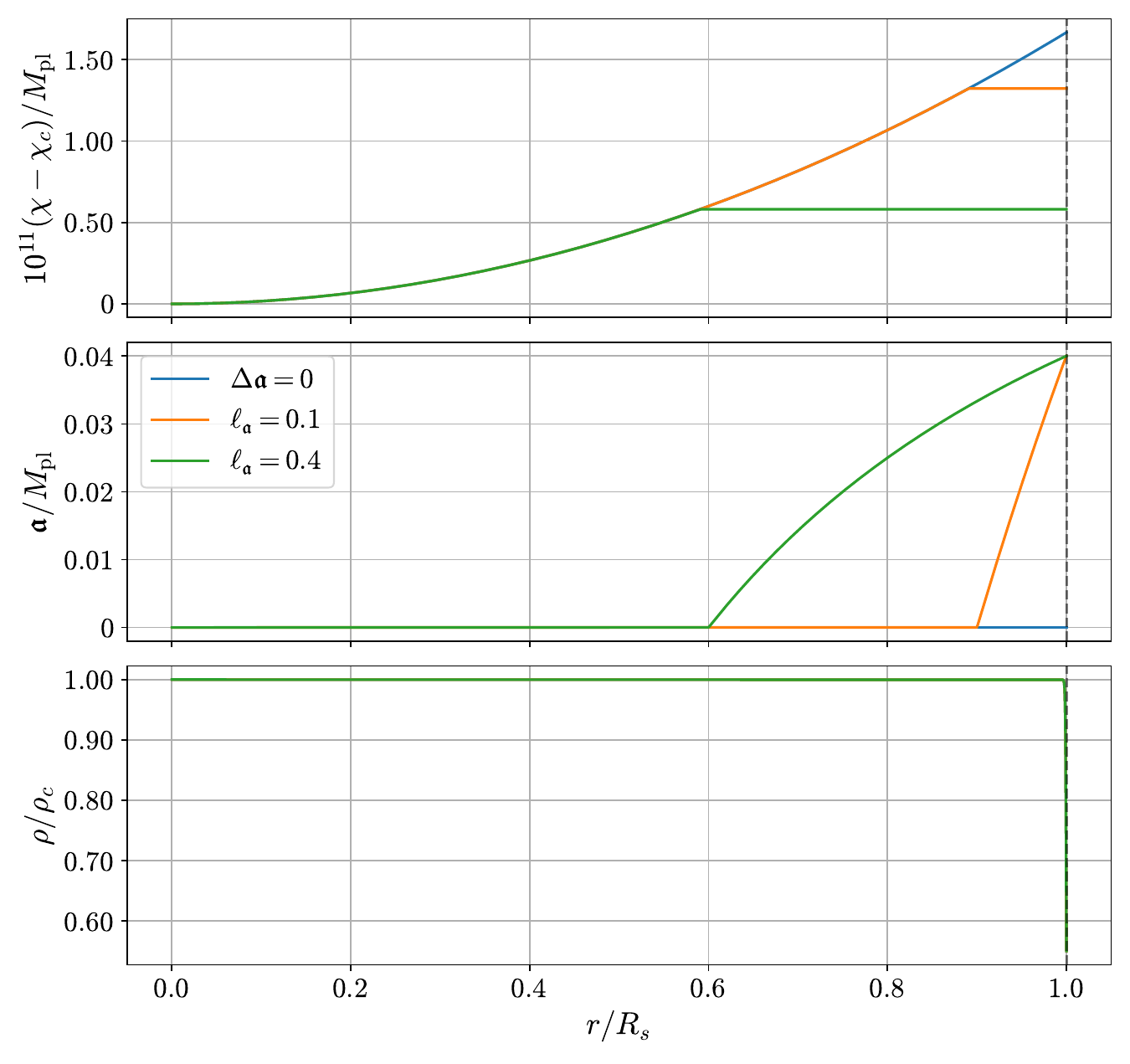}
    \caption{Radial profiles (in units of $r/R_s$) illustrating the
    energy--minimisation (BBQ) regime, for different axion boundary jumps
    $\Delta \mfa=0.04$.  Here we took $W_0 = 1$, $\beta = 0.2$, $\zeta = -2\times10^{10}$. This is using the numerical energy minimisation scheme to determine $\chi_s$, outlined in \Appref{app:methodology}.}
    \label{fig:profiles_stacked_bbq}
\end{figure}

\begin{figure}
    \centering
    \includegraphics[width=\linewidth]{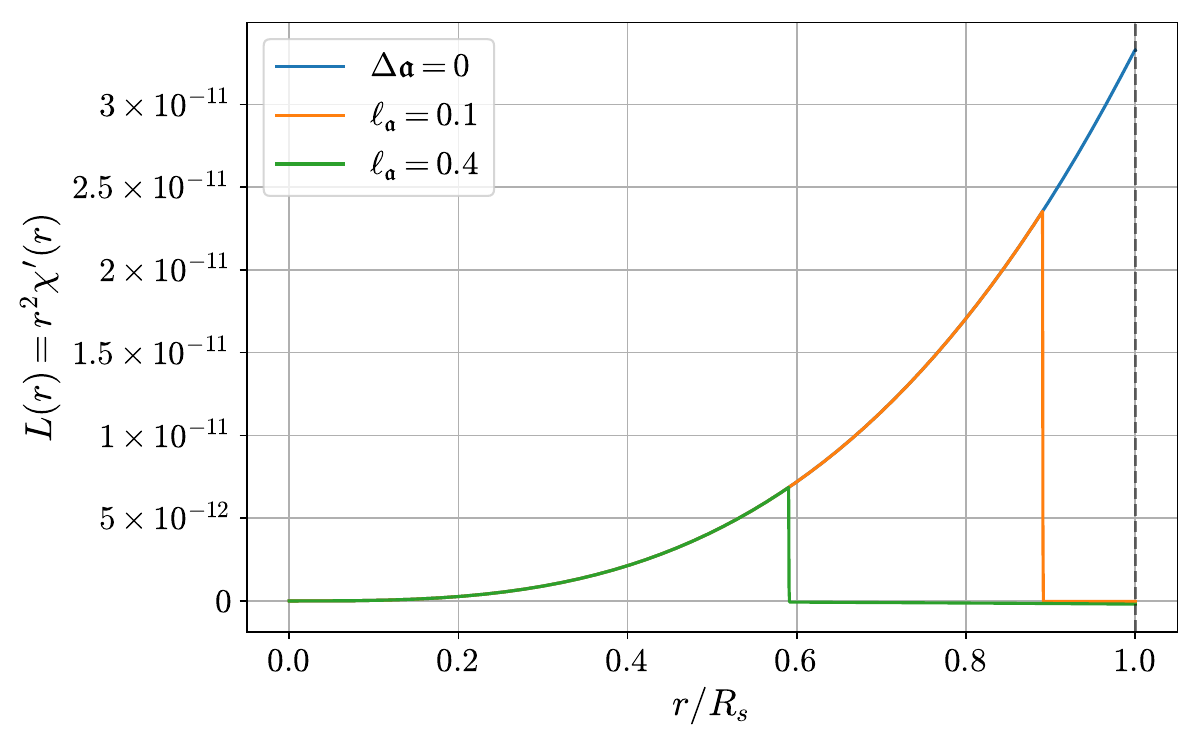}
   \caption{Dilaton charge $L(r)\equiv r^{2}\chi'(r)$ profiles for the BBQ flux solutions
 as a function of radius $r/R_s$ for different axion ramp
widths $\ell_\mfa$. 
The $\Delta\mathfrak{a}=0$ curve reproduces the single-field baseline single field result.
}
    \label{fig:L_bbq}
\end{figure}

This behaviour is cleanly exhibited by the numerical solutions shown in
\Cref{fig:profiles_stacked_bbq} and \Cref{fig:L_bbq} for a constant density object. The dilaton profile evolves throughout the whole interior of the object, consistent with $R_{\rm roll}$
lying deep inside the surface (so $\Delta R=\cO(R_s)$) meaning we are far away from the large environmental mass pinned dilaton regime, while the exterior charge
is nevertheless strongly suppressed once the axion ramp is present.  The BBQ/energy--minimisation mechanism therefore realises
screening in the regime most relevant for a dark--energy like, cosmologically light dilaton, where
$m_{\rm eff}R_s\ll 1$.

These figures also show the efficiency of screening is preserved as one goes beyond the thin axion gradient approximation used in deriving \cref{eq:L-L0-final-clean}.
Physically, the success of screening even with gradual axion gradients is a direct consequence of the large energetic cost associated with the dilaton’s exterior tail. Outside the source,
$\chi' = L/r^{2}$, so the exterior gradient energy scales as
$E_{\rm out}^{(\nabla\chi)} \propto L^{2}/R_s$. The minimum--energy
configuration therefore favours solutions with smaller $L\propto\chi'(R_s)$.

In the BBQ regime the system achieves this by axion backreaction. Because the
axion gradient depends on the local value of $W(\chi)$, the coupled equations
realise a self--regulating feedback, where shifting $\chi$ reshapes the axion contribution to
the dilaton equation. For a sufficiently large hierarchy between the axion
mass inside and outside the source (so that the axion adiabatically tracks its
effective potential; see \Appref{app:methodology}), one finds
$\mfa'(r)\propto 1/W^2(\chi)$. 
In the BBQ regime $W(\chi)$ is monotonic with $W_{,\chi}<0$ over the
relevant field range, so driving the dilaton to smaller $\chi$ increases
$W(\chi)$. In the adiabatic regime the axion gradient is then dynamically
suppressed. This prevents the axion contribution from overshooting the matter
and potential terms in \cref{eq:eom_dilaton}. If it begins to dominate and pushes $\chi$ downwards, the
resulting increase in $W$ reduces $\mfa'$, weakening the axion backreaction and
stabilising the system near the value of $\chi_s$ for which the net sourcing
in the rolling region is minimised. This is why the dilaton profile is ruthlessly flattened and the charge is largely suppressed in \Cref{fig:profiles_stacked_bbq} and \Cref{fig:L_bbq}.

\begin{figure}
    \centering
    \includegraphics[width=\linewidth]{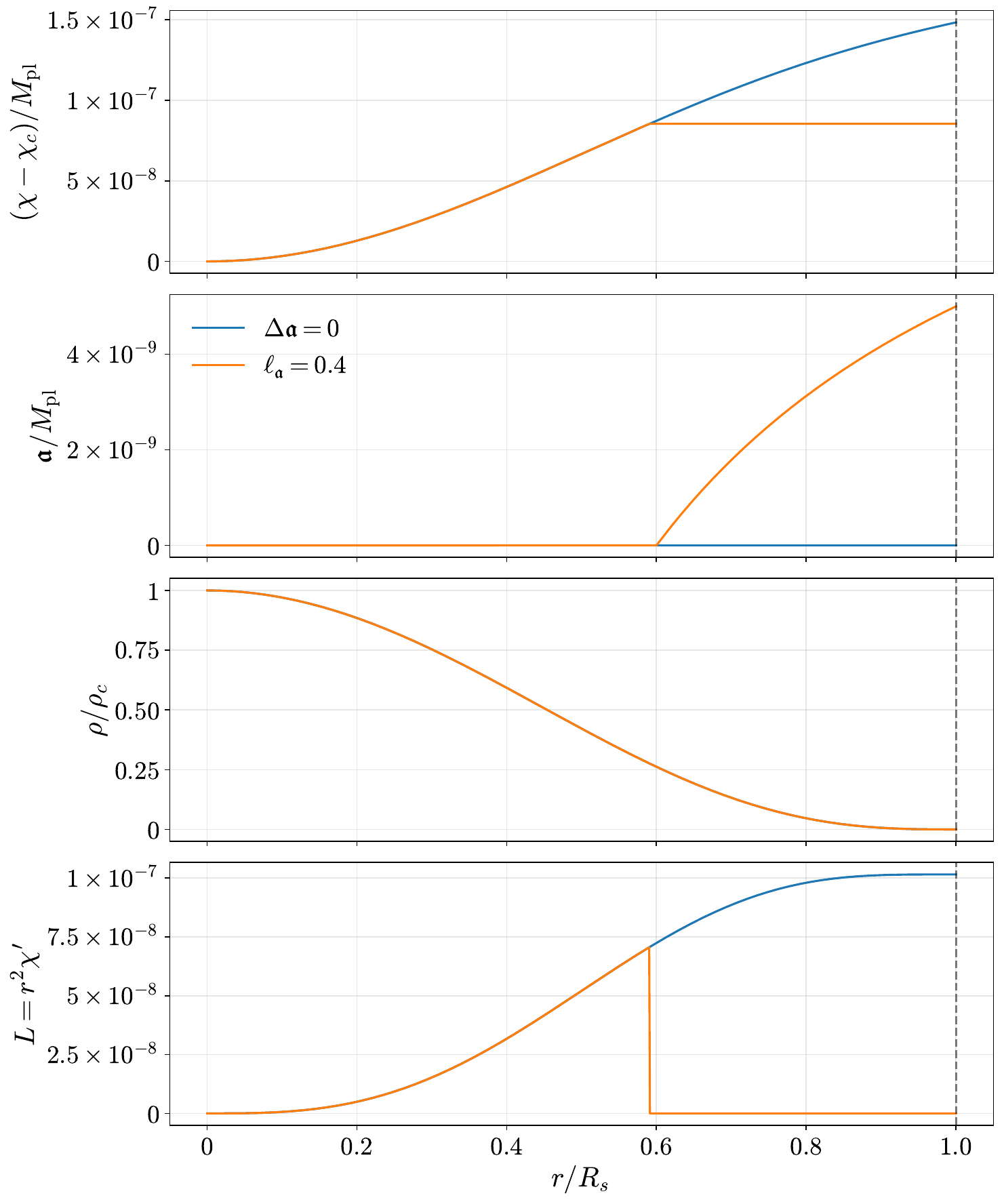}
    \caption{Radial profiles in the presence of a stellar mass density (see \Appref{app:methodology} for details) illustrating the
    energy--minimisation (BBQ) regime, for an axion boundary jump
    $\Delta \mfa=4\times10^{-9}$.  Here we took $W_0 = 1$, $\beta = 0.2$, $\zeta = -2\times10^{10}$. This is using the numerical energy minimisation scheme to determine $\chi_s$, also outlined in \Appref{app:methodology}.}
    \label{fig:profiles_star_bbq}
\end{figure}

\Cref{fig:profiles_star_bbq} shows that the same mechanism operates for a
stellar density distribution, where one might naively expect the screening
efficiency to be degraded. Instead, the dilaton gradient and charge persist up
to the radius at which the axion begins its transition. The efficiency of the
screening therefore depends on the detailed axion dynamics, with broader axion
gradients producing more comprehensive screening. These systems can therefore
exhibit \emph{enhanced} screening efficiency relative to configurations with a
sharp density discontinuity at the surface. Physically, a smooth density
profile allows for a more gradual axion transition throughout the stellar
interior, which in turn screens the dilaton deeper inside the object.

\section{The dilaton revisited}
\label{sec:dilaton_revisited}

The preceding sections show that, in multi--field systems, geometric
localisation of a profile and suppression of the observable fifth
force need not coincide.  In the pinned, two--minima regime the axion backreaction
can drive the dilaton transition into an increasingly thin near--surface layer,
but the conserved exterior charge remains set by the global flux balance.  As a
result, a thinner rolling region is neither necessary nor sufficient for
screening: it can be achieved without reducing the long--range $1/r$ tail, and
can even enhance the charge once the axion--induced contribution to the flux is
accounted for.  Nevertheless, localisation remains phenomenologically relevant
in its own right.  Many laboratory and Solar--System tests probe the detailed field gradients interior to sources, such as through stellar burning and geochemical constraints, so confining
large gradients to a narrow surface layer can reduce the spacetime volume over
which such effects operate even when the asymptotic force is controlled by the
unchanged exterior charge.

By contrast, when the dilaton is ultra--light on astrophysical scales, or when
its effective potential admits no relevant density--dependent minima, the
thin--shell picture ceases to be parametrically well defined.  In this unpinned
regime the dilaton typically rolls throughout the interior, and viable screening
must proceed by suppressing the net scalar charge directly.  In the axio--dilaton
system this occurs through the energy--minimisation (BBQ) mechanism. In this regime the coupled
configuration dynamically selects a surface value that lowers the total static
energy by reducing the exterior gradient energy (by leveraging the axion-dilaton kinetic coupling, with appropriate sign and magnitude to suppress the dilaton's gradient), and hence the conserved charge,
without requiring a geometrically thin rolling layer or an object--by--object
tuning of unrelated contributions.

A remaining model--building question is the size of the kinetic slope required
for efficient charge suppression.  In the exponential ansatz
\cref{eq:exp_W}, the strongest suppression only occurs for large negative $\zeta$,
corresponding to a rapidly varying field--space metric.  Consequently, the
large--$|\zeta|$ regime required for efficient charge suppression should be
interpreted as an \emph{effective} two--field description.  In string
compactifications the function $W(\chi)$ is fixed by the moduli--space metric, so $|\zeta|$ being large is a geometric property of the field space. It requires that the
axion kinetic metric $K_{\mfa\mfa}(\phi^I)$ vary rapidly along the light canonically
normalised direction $\chi$, i.e.
\begin{equation}
|\zeta|\;\sim\;\frac12\left|\frac{d\ln K_{\mfa\mfa}}{d\chi}\right|\gg 1,
\end{equation}
rather than being an arbitrary parameter.  Realising this in a UV completion
therefore amounts to engineering a compactification in which the axionic decay
constant changes sharply as one moves in moduli space along the $\chi$ direction,
for example because the light direction is an aligned/mixed combination of
several moduli so that modest motion in $\chi$ corresponds to a large fractional
change in the underlying geometric modulus controlling the axion metric.
Constructing explicit compactifications that realise such an effectively
large--$|\zeta|$ while maintaining parametric control (without running into the usual disasters of towers of light states
and $\alpha'$ corrections etc.) is therefore an open model--building problem.

Finally, multi--field dynamics can introduce new sources of non--universality
even when the dilaton coupling $A(\chi)$ is universal.  Whenever
the axion sector couples (even weakly) to composition--dependent quantities (such
as baryon number in order to generate the required ramp-like structures interior
to objects that we have shown give screening effects), the density--dependent
equilibrium data that set the axion jump, and hence the effective ramp entering
the dilaton backreaction, become object dependent through $\mfa_\pm$.
This makes the screening of light scalars inherently object- and
composition-dependent, because the same microscopic parameters need not generate the
same axion ramp (or even any ramp) in environments with different matter content,
for example in galaxy haloes where the dominant component is not baryonic.

\section{Conclusions}\label{sec:conclusions}

This work has shown that a constant--coupling dilaton is not generically doomed by fifth--force bounds once its axionic partner is brought into the mix. In an axio--dilaton system the kinetic coupling makes the axion gradients feed directly into the dilaton equation. That contribution enters the flux balance that fixes the conserved exterior charge $L=r^2\chi'$, and therefore the strength of the long--range fifth force. With $W_{,\chi}<0$ the axion backreaction can oppose the matter--induced sourcing of the dilaton and drive $L$ down, providing a concrete route to screening in precisely the regimes where the single--field thin--shell picture fails.

What makes the mechanism work in practice is that the system is not trying to engineer a perfect surface transition of the dilaton to the asymptotic equilibrium. In the unpinned regime, relevant for a cosmologically light dilaton and for models in which the dense--matter minimum is inaccessible on stellar and planetary scales, the configuration is selected by minimising the static energy. The dominant lever in that minimisation is the exterior tail, for which outside the source one has $\chi'=L/r^2$, so the exterior gradient energy scales as $E_{\rm out}^{(\nabla\chi)}\propto L^2/R_s$. Reducing $|L|$ is therefore an efficient way to lower the energy, and the axion provides the channel through which the system can actually do it. Because the axion gradient itself depends on $W(\chi)$, shifting the dilaton reshapes the axion contribution to the flux, which in turn reshapes the dilaton profile. The coupled equations therefore organise a feedback that naturally steers the solution towards the low--charge regime. The effect is that screening survives when the axion ramp is not infinitesimally thin and a finite--width transition still contributes to the same flux balance and still participates in the same energy selection. Our numerical solutions show that the suppression of $L$ persists for broad ramps and for smooth stellar density profiles, in which the axion transition can extend into the interior.

The pinned regime tells the complementary story. When the dilaton is heavy enough to track density--dependent minima, the interior--exterior mismatch already fixes a substantial flux budget. The axion can still act with the opposite sign and partially cancel the charge, but now it is cancelling something that is enforced by pinning rather than something the system is free to relax away. As a result the cancellation is typically delicate and object dependent. The choice of axion gradient amplitude and shape that suppresses the charge for one density profile need not do so for another. This is the sense in which axion gradients can dramatically reshape the near--surface profile without guaranteeing a correspondingly small fifth force.

These results point to several natural extensions. The first is genuinely multi--body configurations. In a realistic solar system or galactic environment there is no reason to expect a single ambient value to be reached independently around each object, and matching the exterior tails of multiple bodies should induce additional gradients in the regions between them. Determining how the effective charges renormalise in that setting requires solving the coupled system with the correct collective boundary conditions. Such an effort is currently underway.

A second direction concerns environmental and composition dependence. Even if the dilaton coupling $A(\chi)$ is universal, the axion sector can introduce object dependence through the mechanism that induces its own gradient to suppress the dilaton's, since these are set by the microphysics that determines how the axion feels matter. The fact that screening does not require an infinitesimal surface jump is important here and it suggests that comparatively modest axion gradients can already be relevant, so the same physics that makes screening robust can also make equivalence--principle sensitive signatures plausible, depending on which matter species source the axion and how.

Thirdly, and complementary to the previous argument, it is known that screening in astrophysical systems behaves very differently once the spherical symmetry assumption is generalised \cite{Burrage:2014daa}. Given that the bodies most sensitive to solar system tests of gravity all roughly obey this spherical symmetry, analysing test bodies that break this assumption, either in the lab or in astrophysical systems can provide a route to detecting modifications to gravity that are otherwise screened \cite{Burrage:2023eol,Almasi:2015zpa,Brax:2007vm,Anson:2020fum}. Understanding how these effects would modify multi-field systems through additional non-linearities could provide detection mechanisms not available to single field tests of modified gravity.

A fourth issue is the model--building meaning of the large negative kinetic slope required for strong suppression in the exponential ansatz $W(\chi)=W_0 e^{\zeta\chi}$. In a UV completion $\zeta$ is a property of the scalar manifold rather than a freely dialled parameter, so the question becomes whether controlled compactifications can realise an effectively large negative $\zeta$ without invalidating the two--field description.

Finally, neutron stars sit in a qualitatively different corner of parameter space. Their densities and relativistic structure can pin a dilaton that is cosmologically light, while the axion dynamics may depend sensitively on the interior composition and on strong gravity. Extending the present analysis to that regime therefore requires a relativistic treatment of the coupled fields in a realistic stellar background, but it also offers the prospect of distinctive signatures and stringent tests of multi--field screening.

More generally, the motivation for studying screening in multi--field
systems extends beyond the specific axio--dilaton model considered here.
Many cosmological scenarios invoke additional gravitational couplings or
new propagating degrees of freedom in order to obtain nontrivial
late--time dynamics, including scalar--tensor and modified--gravity
models capable of producing phantom--divide crossing or other
dark--sector phenomenology
\cite{Perrotta:1999am,Boisseau:2000pr,DeFelice:2009aj,deRham:2010kj,Kimura:2011td,Banijamali:2012zzb,deRham:2012fw,Ip:2015qsa,Crisostomi:2019yfo,Deskins:2019ajp,Iyonaga:2021yfv,Kobayashi:2023lyt,Mylova:2023ddj,Tsujikawa:2025wca,Yao:2025wlx}.
Because the same interactions generically mediate long--range scalar
forces or modify gravity on small scales, compatibility with laboratory
and Solar--System tests requires that their effects be suppressed in
dense environments. Screening is therefore a structural ingredient of
many such cosmological constructions rather than an optional addition.

We hope these considerations motivate renewed interest in the sparsely explored landscape of multi-field astrophysical modelling.

\section*{Acknowledgements}

We thank Cliff Burgess, Sergio Sevillano Muñoz and Maria Mylova for helpful and illuminating discussions. This work grew out of discussions at UK Cosmo 2024 at King’s College London. AS, CvdB, and ACD thank King’s College London for its hospitality during the programme. AS also thanks the Perimeter Institute and Kavli IPMU for their hospitality, where much of this research was carried out. AS is supported by the W.D. Collins Scholarship. CvdB is supported by the Lancaster–Sheffield Consortium for Fundamental Physics under STFC grant: ST/X000621/1. ACD is partially supported by the Science and Technology Facilities Council (STFC) through the STFC consolidated grant ST/T000694/1.

\bibliographystyle{apsrev4-2}
\bibliography{bibliography}

\appendix
\crefname{section}{Appendix}{Appendices}
\Crefname{section}{Appendix}{Appendices}

\section{Quadratic kinetic prefactor}
\label{app:quadraticW}

To complement the exponential kinetic prefactor used in the main text, we briefly
consider a quadratic growth of the axion kinetic function. This provides a simple
example in which $W(\chi)$ increases only polynomially with the dilaton, and it
illustrates how the efficiency and spatial localisation of the axion backreaction
depend on the rate at which the field-space metric grows.

We take
\begin{equation}
  W^2(\chi)
  = 1+\frac{(\chi-\chi_*)^2}{2\Lambda_\chi^2},
  \label{eq:W_quad_def}
\end{equation}
where $\chi_*$ parametrises the location of the minimum in the kinetic sector
and $\Lambda_\chi$ controls the rate of growth.

\begin{figure}
    \centering
    \includegraphics[width=\linewidth]{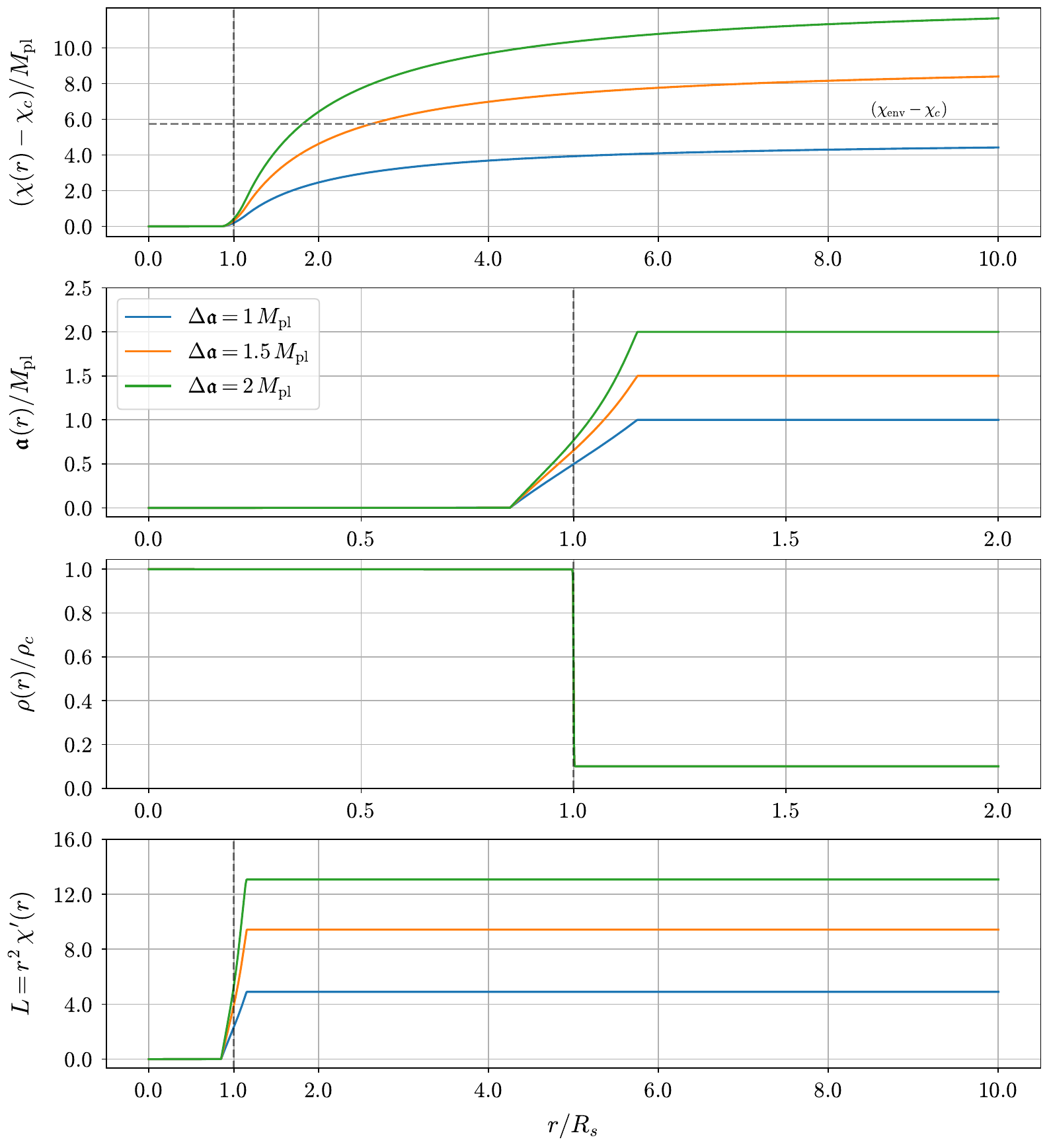}
    \caption{Radial profiles as functions of $r/R_s$ for a quadratic kinetic prefactor
    \cref{eq:W_quad_def}: dilaton $\chi(r)$ (top), axion $\mfa(r)$ (middle), and density
    $\rho(r)$ (bottom), for different axion boundary jumps $\Delta\mfa=\mfa_{+}-\mfa_{-}$.
    We take $\Lambda_\chi=\MPL$ and, for numerical illustration, choose an Earth--like
    density $\rho_c\simeq 10^{-9}\MPL^2/R_s^2$, with $V_0=0.5\,\rho_c$ and
    $\rho_{\rm env}=0.1\,\rho_c$ (an unrealistically small hierarchy used only to make the
    effect visually clear).}
    \label{fig:profiles_quadW}
\end{figure}

The qualitative behaviour of the quadratic--$W$ solutions is shown in
\Cref{fig:profiles_quadW} and is most transparently understood in the
flux--conserving formulation used in the numerical methods outlined in \Appref{app:methodology}. Assuming there is a large hierarchy of the axion masses interior and exterior to the source (as well as the axion being heavy in both environments) so that it tracks the minimum of its effective potential,
in spherical symmetry the axion equation implies a conserved radial flux $J$,
so that
\begin{equation}
\mfa'(r)=\frac{J}{r^2\,W^2(\chi)}\,.
\label{eq:axion_flux_quadraticW}
\end{equation}
For an axion assisted transition one needs $W,_\chi>0$.
As the dilaton evolves to larger values outside matter and $W(\chi)$ increases, the axion gradient is therefore
dynamically quenched according to $\mfa'\propto W^{-2}$, reducing the magnitude
of the axion backreaction term in the dilaton equation \cref{eq:eom_dilaton}.
The extent of the axion forcing is thus controlled by how quickly $W(\chi)$ grows along
the profile.

For an exponential prefactor, $W(\chi)\propto e^{\zeta\chi/\MPL}$, even modest motion of
$\chi$ near the surface produces a rapid increase in $W$, and the axion flux is quenched
efficiently. The axion gradient therefore switches off over a narrow radial interval once
the dilaton begins to roll, confining the external driving effect to a thin region close to
$r\simeq R_s$ and limiting the integrated impact of the axion on the dilaton profile.

By contrast, for the quadratic form \cref{eq:W_quad_def}, the increase of $W(\chi)$ is
only polynomial. The quenching $\mfa'\propto W^{-2}$ is correspondingly milder, so a
substantial axion gradient can persist across a broader part of the surface region while
$\chi$ evolves. This sustains the two--derivative driving over a larger radial interval and
leads to a stronger overall response of $\chi(r)$ than in the exponential case at otherwise
comparable parameters, consistent with the profiles in \Cref{fig:profiles_quadW}. The result is the heightened efficiency of the axion in producing a thin-shell for the dilaton.

\subsection*{BBQ Screening ($W,_\chi<0$)}
\label{app:quadraticW_BBQ}

For completeness we record the analogue of the BBQ energy--minimisation result
when the axion kinetic prefactor is quadratic rather than exponential. This illustrates
the generality of the charge--suppression mechanism for monotonic kinetic sectors
in which $W_{,\chi}$ does not change sign while the growth of $W(\chi)$ is
parametrically milder than exponential.

Repeating the minimisation of the surface energy and exterior gradient
energy in \cref{eq:Etot-compact} yields the surface value
\begin{equation}
  \chi_s - \chi_*
  = \frac{2\beta\,\Phi_N(R_s)\,\MPL}{
         1
         + \frac{R_s}{\ell_\mfa}
           \left(\frac{\Delta\mfa}{\Lambda_\chi}\right)^2
         }\,.
  \label{eq:chi_s_quadratic}
\end{equation}
So the surface value is driven to $\chi_*$ if $\frac{R_s}{\ell_\mfa}\left(\frac{\Delta\mfa}{\Lambda_\chi}\right)^2 \gg 1$. Suggesting the axion transition amplitude $\Delta\mfa$ must be much greater than the kinetic coupling scale $\Lambda_\chi$. This is in contrast to the exponential case, where the shift of the surface value is now
suppressed directly by the magnitude of the axion gradient $\Delta\mfa$.

The corresponding kinetic sourcing term evaluated at the surface is
\begin{equation}
  W(\chi_s)\,W_{,\chi}(\chi_s)
  = -\frac{2\beta\,\Phi_N(R_s)\,\MPL}{
         \Lambda_\chi^2
         + R_s \ell_\mfa
           \left(\frac{\Delta\mfa}{\ell_\mfa}\right)^2
       }\,,
  \label{eq:WWchi_quadratic}
\end{equation}
and the exterior scalar charge is therefore suppressed according to
\begin{equation}
    \frac{L}{L_0}
    = \frac{1}{
      1+\frac{R_s \ell_\mfa}{\Lambda_\chi^2}
      \left(\frac{\Delta\mfa}{\ell_\mfa}\right)^2 }\,,
  \label{eq:L_over_L0_quadratic}
\end{equation}
showing explicitly that, for a quadratic kinetic prefactor, increasing the axion
gradient decreases the fifth force monotonically.

\begin{figure}
    \centering
    \includegraphics[width=\linewidth]{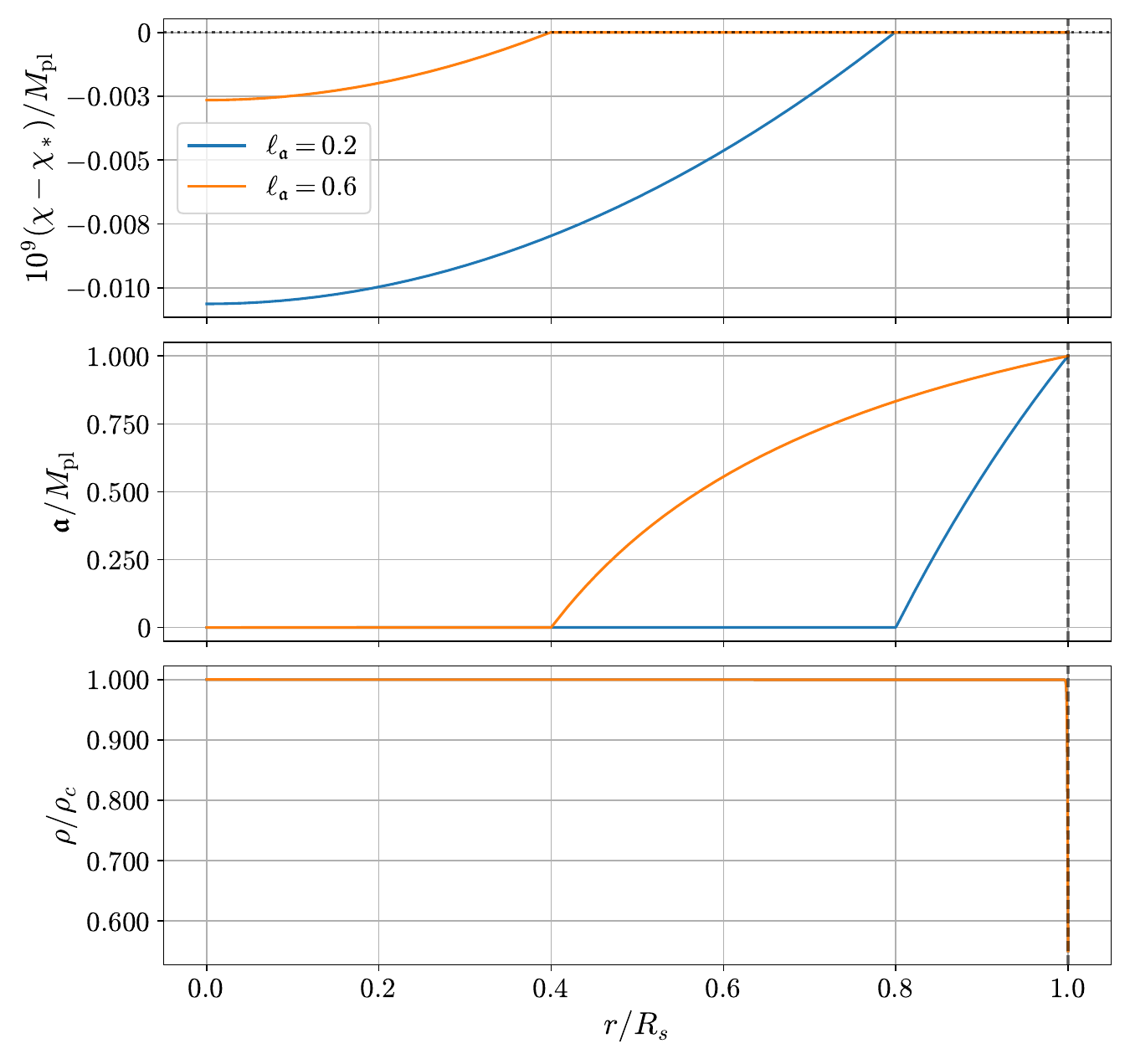}
    \caption{Field profiles interior of a constant density object with quadratic kinetic coupling \cref{eq:W_quad_def} between the axion and dilaton, and $\rho_c$ corresponding to the surface density of the Earth, $V_0 = 0.5\rho_c$ and $\rho_{\rm env} = 0.1\rho_c$. Here we used the quadratic kinetic $\Lambda_\chi = 10^{-5}\MPL$, $\chi_* = 10^{-9}$ and the flux conserving treatment for the axion.}
    \label{fig:bbq_profiles_quad_w}
\end{figure}

\begin{figure}
    \centering
    \includegraphics[width=\linewidth]{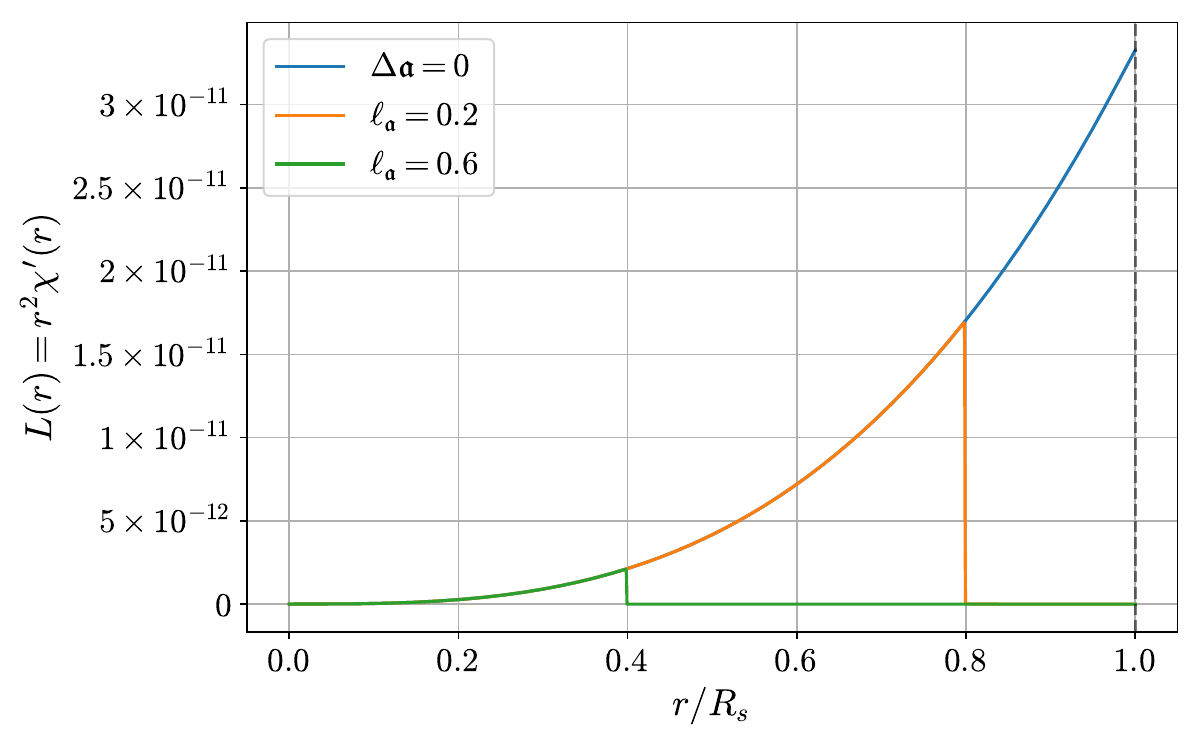}
    \caption{Corresponding charge for the respective cases in \Cref{fig:bbq_profiles_quad_w} using a quadratic kinetic coupling \cref{eq:W_quad_def} between the axion and dilaton.}
    \label{fig:bbq_charge_quad_w}
\end{figure}

Taken together, \cref{eq:chi_s_quadratic,eq:L_over_L0_quadratic} highlight a key
qualitative difference between polynomial and exponential kinetic couplings. For
quadratic $W(\chi)$ the charge suppression depends directly on the axion gradient
and increases smoothly as $\Delta\mfa/\ell_\mfa$ is made larger. This is confirmed in \Cref{fig:bbq_profiles_quad_w} and \Cref{fig:bbq_charge_quad_w} where we can see the dilaton is very quickly driven to $\chi_*$ when the axion gradient is turned on, and held there continuously to the surfaces, suppressing the dilaton's surface charge and its effective coupling to matter.

\section{Numerical methodology}\label{app:methodology}

To test the analytic thin--shell estimates and to explore the fully coupled
axio--dilaton system beyond the analytic approximations of Secs.~III--V, we perform
direct numerical integrations of the spherically symmetric field equations.
The numerics retain the essential physics of the thin--shell mechanism while
introducing controlled approximations that render the problem tractable on a
one–dimensional radial grid.  The code implementing the methods described here
has been made publicly available at \cite{SmithCode2026}.

\subsection*{Overview of numerical schemes}
We employ two complementary numerical strategies for the coupled axio--dilaton
profiles, depending primarily on the regularity of the matter profile at
the surface and the asymptotic values of the dilaton field.

For smooth stellar density profiles $\rho_m(r)$, we solve the full coupled two--field Klein--Gordon equations
directly, imposing regularity at
the origin and an additional appropriate boundary condition dependent on the scenario we study.  This direct two--field
solution is used for \Cref{fig:star_thick_ramp_profiles,fig:profiles_stacked_small_mass_thin_shell,fig:star_pinned_dilaton_gradient_suppression,fig:environmental_profiles,fig:environmental_L}.

In contrast, for an idealised near--discontinuous top--hat density
transition, the coupled Klein--Gordon system becomes
numerically stiff at the surface and develops spurious large gradients localised
to the jump. If not compensated by a regulating asymptotic boundary condition, this can lead to runaway behaviour due to numerical instabilities. Additionally, for simulations such as BBQ energy minimisation, which are computationally intensive, solving the full two-field system becomes untractable numerically. In these cases we avoid solving the axion second--order
equation across the discontinuity and instead reconstruct the axion gradient from
the conserved canonical flux, treating the surface layer as a localised source
term and iterating self--consistently with the dilaton profile.  This flux
reconstruction scheme is the one used for \Cref{fig:thick_ramp_profiles,fig:profiles_stacked_bbq,fig:L_bbq} and we now outline the flux formulation details.

\subsection*{Axion flux reconstruction}

The full radial axion equation reads
\begin{equation}
  \frac{1}{r^2}\frac{d}{dr}\!\left(r^2W^2(\chi)\,\mfa'(r)\right)
  = \mathcal{J}(r),
  \label{eq:axion_full_eom_app}
\end{equation}
where $\mathcal{J}(r)$ encodes the microscopic source driving $\mfa_- \to \mfa_+$.  
Away from this region, the absence of external forces implies the exact conservation law
\begin{equation}
  J \equiv r^2 W^2(\chi)\mfa'(r)
  = \text{constant}.
  \label{eq:flux_conservation_app}
\end{equation}
To remain agnostic about the microphysical form of $\mathcal{J}(r)$, we approximate
the source region by a normalised profile $S(r)$, so that
\begin{equation}
   \frac{1}{r^2}\frac{d}{dr}\!\left(r^2 W^2(\chi)\mfa'\right)
   = J\,S(r).
   \label{eq:axion_eom_S_app}
\end{equation}
The normalisation condition ensures that the total axion excursion is fixed to
$\mfa_+-\mfa_-$, and flux conservation is in-practice achieved by evaluating $J$ as
\begin{equation}\label{eq:normalisation integral}
  \left(\mfa_+ - \mfa_-\right)\int dr\,\frac{S(r)}{r^2 W^2(\chi)} = J^{-1}.
\end{equation}

Throughout this work we adopt a rectangular window
\begin{equation}\label{eq:source_window}
  S(r)=
  \begin{cases}
     1, & \alpha R_s<r<R_s,\\
     0, & \text{otherwise},
  \end{cases}
\end{equation}
so that the interpolation occurs in a thin layer near the surface while the
interior and exterior remain frozen at $\mfa_-$ and $\mfa_+$, respectively.  The
dilaton–axion backreaction is treated exactly through the flux law by \cref{eq:flux_conservation_app}, so that
wherever $W(\chi)$ grows across the shell, the axion gradient $\mfa'$ is
suppressed, stiffening the effective ramp and reproducing the behaviour
dictated by the full equations of motion.

\subsubsection*{Consistency with full solutions}

Figure~\ref{fig:bbq_charge_quad_w} provides a direct numerical check that the
flux--conservation treatment of the axion reproduces the
profiles obtained from the full coupled boundary--value problem, without
altering the underlying microphysical axion sector.
The full reference solution
(solid curves) is obtained with the stellar--profile solver described in
\Cref{sec:hybrid_solver}, where the axion dynamics is governed by the same microphysical
potential and matter coupling as in the main text, \cref{eq:V_ax,eq:U_ax}.

To assess the flux formulation, we then perform an additional run in
which the axion equation is replaced by a conserved flux ansatz localised to a
narrow window around the surface, implemented through the window function
$S(r)$ supported on $r\in(R_c-\ell_a/2,\,R_c+\ell_a/2)$.
Crucially, the code does not change the microphysical parameters
$\mfa_\pm$ appearing in \cref{eq:V_ax,eq:U_ax}. In particular,
$\mfa_-$ is left untouched, so the same $U(\mfa)$ is used in both
the full and flux runs.
Instead, the flux normalisation $J$ is fixed \emph{solely} by matching the
endpoint values of the \emph{full} axion solution,
\begin{equation}
\mfa_{\rm flux}(r_{\min})=\mfa_{\rm full}(r_{\min}),\qquad
\mfa_{\rm flux}(r_{\max})=\mfa_{\rm full}(r_{\max}).
\end{equation}
Operationally, this determines $J$ from the integral constraint implied by the
windowed flux, while leaving $\mfa_\pm$ (and hence \cref{eq:V_ax,eq:U_ax})
unchanged.

\begin{figure}
    \centering
    \includegraphics[width=\linewidth]{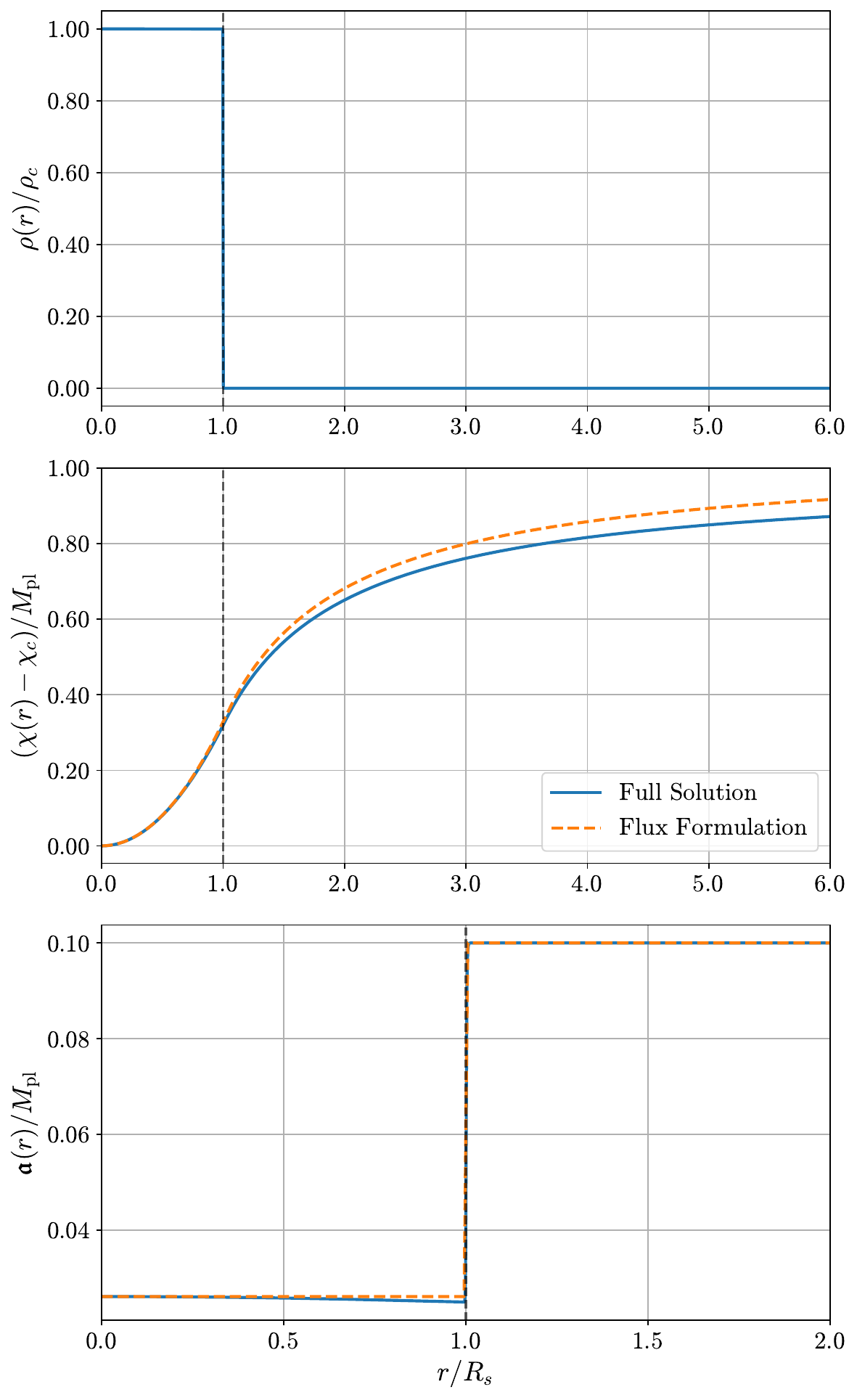}
    \caption{Radial profiles for the coupled axion--dilaton system in a smoothed top-hat halo. \textbf{Top:} density contrast $\rho(r)/\rho_c$. \textbf{Middle:} dilaton displacement $(\chi(r)-\chi_c)/M_{\rm pl}$ for the full coupled solution (solid) and the flux-conservation axion formulation (dashed), with the flux boundary values matched to the full solution endpoints. \textbf{Bottom:} axion field $\mathfrak{a}(r)/M_{\rm pl}$ for the same two runs.}
    \label{fig:flux_vs_full}
\end{figure}

With this setup, any difference between the solid (full) and dashed (flux)
curves in Fig.~\ref{fig:flux_vs_full} is attributable to the flux ansatz
itself rather than to retuning the axion microphysics.
The close agreement observed therefore validates the flux--conservation
approximation as a faithful representation of the full coupled solution in the
regime considered.

\subsection*{Matter density profile}

Here we give the explicit functional forms of the density distributions used in the numerics throughout the paper.

\subsubsection*{Idealised planetary mass distribution}
The source is modelled as a sphere of radius $R_s$ with central density $\rho_c$
embedded in an environment of density $\rho_{\rm env}$. 
The central density is taken illustratively as that of the Earth's so that $\Phi_N(R_s) \sim 10^{-9}$.
To avoid an unphysical
discontinuity in the force at the surface we introduce a smooth transition
\begin{align}\label{eq:top_hat_ish}
  \rho_m(r)
  = \tfrac12 &\rho_c\!\left[1 - \tanh\!\left(\frac{r-R_s}{\ell_\rho}\right)\right]
  \nn\\&\qquad\qquad\qquad+ \tfrac12 \rho_{\rm env}\!\left[1 + \tanh\!\left(\frac{r-R_s}{\ell_\rho}\right)\right],
\end{align}
with $\ell_\rho\ll R_s$ being the characteristic width of the surface density gradient.  This retains the essential density contrast while
remaining differentiable.

\subsubsection*{Stellar mass density distribution}
For stellar applications we adopt a smoothened truncated polytropic profile
(used as a simple analytic proxy for a Lane--Emden solution)\cite{chandrasekhar, KippenhahnWeigertWeiss2012}.  The density profile of a star falls approximately as
\begin{equation}\label{eq:stellar_dist}
  \rho_{\rm poly}(r)
  = \rho_c\left[1-\left(\frac{r}{R_s}\right)^2\right]^{n_{\rm poly}},
  \qquad r\le R_s,
\end{equation}
while outside we add an additional environmental density $\rho_{\rm env}$ for $r>R_s$. The exponent $n_{\rm poly}$ controls the
steepness of the stellar stratification (with $n_{\rm poly}=3$ used as a
representative choice in our numerics). The core density is taken to ensure that the surface Newtonian potential is given by $\Phi_N(R_s)\sim 10^{-6}$, which is that of the Sun.

\subsection*{Multi-field solver formulations}

Given the different case studies considered in this paper and the differing appropriate numerical schemes for each one, we outline here the three types of solvers that were made use of for the numerical simulations of this paper, along with the corresponding plots that were produced with each.

\subsubsection*{IVP formulation in the runaway case}

For sharply localised surface features (in particular the near--top--hat profiles used in
\cref{eq:top_hat_ish}), the coupled system can become numerically stiff in the vicinity
of $r\simeq R_s$ and can easily enter into non-physical runaway regimes.  In these runs the dilaton does not reliably relax to a well--defined exterior
minimum within the numerical domain, so there is no stable asymptotic condition that would
select a unique boundary solution for $\chi$.  We therefore evolve the dilaton equation as an
initial--value problem (IVP), imposing regularity at a small inner radius $r_{\rm in}\ll R_s$,
\begin{equation}
  \chi'(0)=0,
  \qquad
  \chi(0)=\chi_c,
\end{equation}
and integrating outward to $r_{\rm out}\gg R_s$ with a Runge-Kutta solver.  When a local minimum
of $V_{\rm eff}$ exists at the core density, we take $\chi_c$ to be that minimum.

The axion is not evolved as a second--order IVP in this regime.  Instead, we fix the net
excursion $\Delta\mfa=\mfa_+-\mfa_-$ and reconstruct $\mfa'(r)$ from the conserved flux relation
\cref{eq:flux_conservation_app} together with the chosen source window $S(r)$, as described in
\cref{eq:source_window}.  Since the reconstructed $\mfa'(r)$ depends on $\chi(r)$
through $W(\chi)$, while the dilaton equation depends on $\mfa'(r)$ through the backreaction
term, the profiles are obtained by a Picard iteration: given a current axion
profile (equivalently a current $J$), we integrate the dilaton IVP to obtain $\chi(r)$, update
$J$ using the normalisation condition \cref{eq:normalisation integral}, reconstruct $\mfa'(r)$,
and repeat until both $J$ and $\chi(r)$ change by less than fixed tolerances. 

This is the integration scheme used for figure \Cref{fig:thick_ramp_profiles}.

\subsubsection*{Hybrid Axion--BVP, Dilaton--IVP solver}
\label{sec:hybrid_solver}

The pure IVP approach above is tailored to near--discontinuous surface features, for which
no reliable asymptotic selection is available within a finite numerical domain.  For the
smooth stellar profiles considered in \cref{eq:stellar_dist}, the exterior evolution is
well behaved and the physically relevant branch of the axion field is most cleanly imposed
by boundary conditions at large radius.  In this regime we therefore adopt a hybrid
scheme.  We evolve the dilaton as an IVP from the regular interior, while solving the axion
as a boundary-value problem (BVP) to enforce the desired asymptotic behaviour.  The two fields are coupled by a
Picard iteration.

At a given iteration, we first hold $\chi(r)$ fixed and solve the full second--order axion
equation as a boundary--value problem, supplying the $\chi$--dependence of the coefficients
through interpolation of the current dilaton iterate (including $\chi'$ where required).
We impose regularity at the origin and fix the exterior branch by the asymptotic value
\begin{equation}
  \mfa'(0)=0,
  \qquad
  \mfa(r_{\max})=\mfa_+,
\end{equation}
with $r_{\max}\gg R_s$ chosen large enough that the solution has reached its asymptotic
regime.  With the updated axion profile held fixed, we then integrate the dilaton equation
outward as an IVP with interior initial conditions
\begin{equation}
  \chi(0)=\chi_c,
  \qquad
  \chi'(0)=0,
\end{equation}
where $\chi_c$ is determined from the local minimum condition of $V_{\rm eff}$ at the core
density.  To resolve the surface region efficiently, the IVP is evolved with a reduced
maximum step size in a band around $r\simeq R_s$.

Starting from smooth initial guesses for $(\chi,\mfa)$, we alternate the axion--BVP update
and the dilaton--IVP update until both profiles stabilise.  The axion update is applied in
Gauss--Seidel fashion so that the axion boundary conditions are satisfied at every iteration.
The dilaton update is damped,
\begin{equation}
  \chi^{(n+1)}=\chi^{(n)}+\omega_\chi\!\left[\chi_{\rm IVP}^{(n)}-\chi^{(n)}\right],
\end{equation}
with $\omega_\chi$ chosen to control large changes and improve stability.
Convergence is declared when the maximum changes with iterations in both $\chi$ and $\mfa$ fall
below fixed tolerances.

For large excursions $\Delta\mfa$ the coupled iteration can become sensitive to the initial
guess.  In that case we employ a homotopy in the outer boundary value by solving a sequence
of problems with $\mfa_+$ increased in steps, warm--starting each run from the converged
solution at the previous $\mfa_+$ (both for the axion BVP and for the interpolated $\chi,\chi'$
entering its coefficients).  This is the scheme used for
\Cref{fig:star_thick_ramp_profiles,fig:star_pinned_dilaton_gradient_suppression}.  The environmental
profiles in \Cref{fig:environmental_profiles,fig:environmental_L} are obtained by wrapping the same
procedure inside a bounded scan over $\zeta$, holding all other parameters fixed.

\subsubsection*{Boundary--value formulation when $V_{\rm eff}$ has minima}

If $V_{\rm eff}(\chi)$ admits a density--dependent asymptotic minimum outside the matter
distribution, the dilaton profile can be obtained by solving the second--order
radial equation with two boundary conditions as a BVP.  Regularity at the origin always
enforces
\begin{equation}
  \chi'(0)=0.
\end{equation}
The second boundary condition is obtained by imposing the asymptotic condition
\begin{equation}
  \chi(r_{\max})=\chi_{\rm env},
\end{equation}
at a large radius $r_{\max}\gg R_s$, where $\chi_{\rm env}$ is the exterior minimum
of $V_{\rm eff}$, and integrate inward or outward accordingly.

In this regime the axion can likewise be solved as a BVP by taking
$\mfa'(0)=0$, ensuring regularity at the origin, while the asymptotic boundary
condition $\mfa(r_{\rm max})=\mfa_+$ to select the correct physical transition.  This
procedure yields a unique pair $(\chi,\mfa)$ that satisfies the full two--field
equations, and integration methods converge reliably. This is used in the modelling for \Cref{fig:profiles_stacked_small_mass_thin_shell}.

\subsection{BBQ Screening}

For the unpinned BBQ configurations of \Cref{fig:profiles_stacked_bbq,fig:L_bbq,fig:profiles_star_bbq} we do not solve the axion as a second--order BVP. Instead,
we adopt the flux--reconstruction setup introduced above.
This choice is because selecting the physical configuration requires an outer minimisation
over the surface value $\chi_s$, which entails rerunning the solver many times. Solving
the full coupled two--field problem at each objective evaluation is substantially more
expensive, whereas the single--field dilaton BVP with a conserved--flux treatment of the
axion captures the relevant backreaction at greatly reduced cost. For the scope of this
work we therefore use this reduced description in the BBQ parameter scans.

For a given trial surface value $\chi_s\equiv \chi(R_s)$ we solve the dilaton profile
inside the object as a boundary--value problem on $[r_{\min},R_s]$, enforcing regularity by imposing
\begin{equation}
  \chi'(0)=0,
  \qquad
  \chi(R_s)=\chi_s.
\end{equation}
The axion gradient is reconstructed from the conserved flux using
\cref{eq:flux_conservation_app} together with the window \cref{eq:source_window}, with the flux
$J$ determined self--consistently by the normalisation condition \cref{eq:normalisation integral}.
Since $J$ depends on $\chi$ through \cref{eq:normalisation integral} while $\chi$ depends on $J$
through the backreaction, the coupled solution at fixed $\chi_s$ is obtained by the same
damped fixed--point iteration described in the preceding subsection.

The physical configuration is then selected by minimising the total energy as a function
of the surface value $\chi_s$. For each $\chi_s$ we evaluate
\begin{align}
  E_{\rm full}(\chi_s)=4\pi&\int_{0}^{R_s}\!dr\,r^2
  \biggl[\tfrac12(\chi')^2+\tfrac12 W^2(\chi)(\mfa')^2\nn\\&\qquad\qquad+V(\chi)+A(\chi)\rho(r)\biggr]
  + \frac{2\pi}{R_s}\,L^2,
\end{align}
where the last term accounts for the exterior tail with constant $L$.
We locate the minimum by a coarse scan in $\chi_s$ over a fixed bracket followed by
bounded local refinements around the best scan candidates, and record the full set of
objective evaluations for diagnostic plots of $E_{\rm full}(\chi_s)$.


\end{document}